%% file: main.tex
\documentclass[10pt, sigplan, nonacm]{acmart}

\AtBeginDocument{%
  \providecommand\BibTeX{{%
    \normalfont B\kern-0.5em{\scshape i\kern-0.25em b}\kern-0.8em\TeX}}}

\usepackage{tabularx}
\usepackage{subcaption}
\usepackage{algorithm}
\usepackage{dsfont}
\usepackage[noend]{algpseudocode}
\usepackage{makecell}
\usepackage{cleveref}
\usepackage{siunitx}
\usepackage{multirow}
\usepackage{pdfrender}
\usepackage{tablefootnote}
\usepackage{natbib}
\usepackage{enumitem}
\usepackage{layout}

\makeatletter
\newcommand\thefontsize{The current font size is: \f@size pt}
\newcommand\thefont{The current font is: \expandafter\string\the\font}

\newcommand{\system}{{WRIST}}
\newcommand{\dataset}{{SPREAD}}
\newcommand{\smodelold}{{SYL-3}}
\newcommand{\smodelnew}{{SYL-4}}

\newcommand{\ignore}[1]{{}}
\newcommand{\bfpara}[1]{\vskip 1ex \noindent \textbf{#1.}}
\newcommand*{\boldcheckmark}{%
  \textpdfrender{
    TextRenderingMode=FillStroke,
    LineWidth=.5pt, 
  }{\checkmark}%
}

\begin{document}

\title{Spectro-Temporal RF Identification using Deep Learning}


\author{Hai N. Nguyen, Marinos Vomvas, Triet Vo-Huu, Guevara Noubir}
\email{{nguyen.hai, m.vomvas, vohuu.t, g.noubir}@northeastern.edu}
\affiliation{%
  \institution{Cybersecurity and Privacy Institute}
  \institution{Northeastern University}
}



\input{abstract}
\maketitle





\input{introduction}


\input{CNN}
\input{synthetic}

\input{real-time}

\input{dataset}

\input{related}

\input{discussion}

\input{conclusion}



\bibliographystyle{ACM-Reference-Format}
\bibliography{main}


\end{document}

%% file: abstract.tex
\begin{abstract}

\ignore{Understanding RF emissions is becoming critical for several security and privacy applications. Wireless softwarization and low-cost SDRs are expanding the attack surface of wireless systems. For instance, smart jammers and weaponized drone attacks are becoming easier to accomplish.} 

RF emissions' detection, classification, and spectro-temporal localization are crucial not only for tasks relating to understanding, managing, and protecting the RF spectrum, but also for safety and security applications such as detecting intruding drones or jammers. Achieving this goal for wideband spectrum and in real-time performance is a challenging problem.
We present \system{}, a \textbf{W}ideband, \textbf{R}eal-time RF \textbf{I}dentification system with \textbf{S}pectro-\textbf{T}emporal detection, framework and system. Our resulting deep learning model is capable to detect, classify, and precisely locate RF emissions in time and frequency using RF samples of 100 MHz spectrum in real-time (over 6Gbps incoming I\&Q streams). Such capabilities are made feasible by leveraging a deep learning-based \textit{one-stage object detection} framework, and transfer learning to a multi-channel image-based RF signals representation. We also introduce an iterative training approach which leverages synthesized and augmented RF data to efficiently build large labelled datasets of RF emissions (\dataset{}). \system{}'s detector achieves 90 mean Average Precision even\ignore{for high Intersection over Union (IoU)} in extremely congested environment in the wild. \system{} model classifies five technologies (Bluetooth, Lightbridge, Wi-Fi, XPD, and ZigBee) and is easily extendable to others. We are making our curated and annotated dataset available to the whole community. It consists of nearly 1 million fully-labelled RF emissions collected from various off-the-shelf wireless radios in a range of environments and spanning the five classes of emissions. 
\end{abstract}

%% file: introduction.tex
\section{Introduction}



Mobile technologies, fueled by advances in wireless communications, revolutionized our society beyond the pioneers dreams. It enables a ubiquitous access to information, and connects people to each other, and to a rapidly increasing number of services. However, a plethora of emerging applications, such as Massive IoT (MIoT), autonomous cars, robotics, and augmented reality are driving the demand for spectrum to new heights. Spectrum scarcity is becoming a critical issue. At the same time, wireless systems are increasingly softwarized, and SDR platforms are highly capable, with small form factor and low cost. For instance, the XTRX SDR platform is capable of 2x2 120MSps in a mini PCIe form factor and costs few hundreds of dollars~\cite{XTRX}. This is both a \textit{blessing} for developing new sophisticated communications techniques (that are agile and flexible, exploiting every pocket of the spectrum), and a \textit{curse} as it calls for new mechanisms for spectrum management and it lowered the barrier for attacks from smart jammers, \ignore{spoofers~\cite{spoof_TPDS13} and eavesdroppers~\cite{ghostbuster_mbcom18}, }to \ignore{misbehaving radios~\cite{IslamLXYLK2018}, }compromised wireless chips~\cite{google-projectzero-wifi}, or weaponizing drones~\cite{isis-drones-nyt,isis-dronesshield}\ignore{ which are increasingly robust and efficient to control~\cite{reactive_msys16}}. While the DHS, FAA, and FCC have regulations against such threats~\cite{FCC,FCC-tip-line,FCC-fine,FCC-Education-Enforcement,FAA-drone,DHS-drone}, they unfortunately, still lack the necessary technology to enforce them. This confluence of trends raises challenging research questions as to the development of scalable techniques for understanding, managing, and protecting the RF spectrum. Some of the traditional areas that will benefit from such techniques include spectrum management, as dynamic and fine-grain spectrum sharing is becoming a necessity even for 5G cellular systems~\cite{GSMA-spectrum-sharing-2019, GSMA2018}. Crucial to all these applications is the ability to understand the spectrum, in \textit{real-time} and \textit{a-posteriori}, detect, classify, and predict communication patterns in time, frequency, and space. Basic spectrum sensing techniques are insufficient as they cannot classify emissions, detect collisions, and adequately summarize the view of wideband spectrum. \ignore{This need has also been recognized by  several initiatives such  as DARPA SC2~\cite{DARPA-SC2}, DARPA RFML~\cite{DARPA-RFML}, and NSF MLWiNS~\cite{NSF-MLWINS} programs.} 

\begin{figure}[t]
    \centering
    \includegraphics[height=2.5cm]{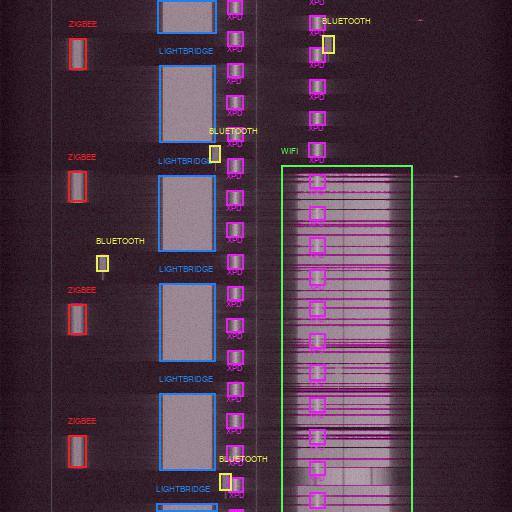}
    \hspace{5mm}
    \includegraphics[height=2.8cm]{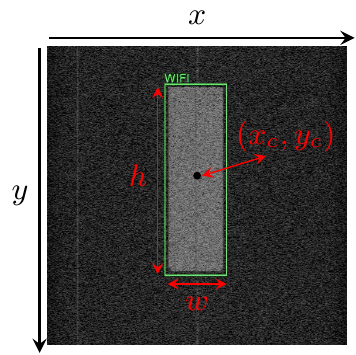}
    \caption{RF emissions are identified as a bounding box located at center frequency $x_c$ and time $y_c$.}
    \label{fig:detection}
\end{figure}

\ignore{Understanding RF emissions, in real time, is a critical building block not only to secure the RF spectrum, but also to understand the privacy threats of devices tracking.} 
We propose \textit{systematic and generalizable approaches} to detect and classify RF emissions with two key unmet requirements: \textit{real-time} and \textit{wideband} spectrum processing. To the best of our knowledge, previous work only focused on a subset of these objectives. We systematically develop RF-Centric ML models and techniques to detect and classify a wide variety of existing wireless standards, to be easily extensible to new and unknown RF emissions. Our approach is inspired by  the success achieved by computer vision in several ways. For the \textit{real-time} spectro-temporal detection and classification of RF emissions, our approach is inspired by YOLO~\cite{yolov1,yolov2,yolov3}. \ignore{YOLO revolutionized computer vision over the last few years, simultaneously achieving real-time detection and classification of objects (into twenty classes), with a high mean average precision (mAP).} In this paper, we generalize and extend the principles underlying YOLO's success to the RF domain. These include (1) analyzing a multi-channel image representation of RF emissions in a single run (unlike prior work that iterates through sliding and resizing, or complex multi-stage pipelines) by creating a grid and detecting/classifying objects per cell, (2) direct location prediction combined with a small number of bounding boxes bootstrapped with anchor training and specialized to learn typical patterns, and (3) fine-grain features detection through passthrough network design and multiscaling.

We believe that developing large curated and labelled datasets, and sharing them with the wireless community will spur the creation of new RFML models and techniques. Towards this goal we developed a dataset of over 1.4 TBytes of RF samples and images including emissions from a variety of radios that operate in the 2.4GHz ISM band including Wi-Fi, Bluetooth, ZigBee, Lightbridge, XPD. The dataset consists of I\&Q samples recorded at 100 MHz. The recordings are structured and annotated in time and frequency with corresponding radio technology classes. The dataset is complemented with tools for printing and parsing recordings, as well as creating auto-labelled synthetic data (\Cref{sec:dataset}).

Towards building the deep learning models, and in the absence of an initial labelled dataset, we developed a set of techniques to minimize manual efforts. We first reused some of YOLO existing layers and weights (transfer learning). On the other hand, we developed an approach to bootstrap an iterative process of using synthetic intermediate data, building increasingly large datasets and accurate deep learning models. 
Our contributions can be summarized as follows:

\begin{itemize}[wide]
    \item An extensible deep learning framework for real-time RF identification inspired by and leveraging transfer learning from state-of-the-art computer vision, and an image-based RF representation to enable such learning~(\Cref{sec:cnn}).
    
    \item An efficient iterative learning approach to develop ML models  consisting of two stages: (1) Transfer learning from a dataset of synthesized and augmented RF data, and (2) Re-learning from a large dataset of over-the-air RF emissions acquired using the previous model (\Cref{sec:online,sec:offline}).
    
    \item Deep learning architecture and models for real-time, wideband RF emissions analysis, detection, localization, and classification achieving 90 mean Average Precision (mAP) even in extremely congested environment in the wild.  
    
    \item Collecting, curating, and publishing a 1.4 TBytes dataset (\dataset{}) of nearly 1 million fully-labelled RF emissions (also submitted as a MobiSys Artifact). We also introduce efficient and systematic approaches to extend the dataset to new waveforms providing a building block for RFML and wireless networking research (\Cref{sec:dataset}).

\end{itemize}

%% file: CNN.tex
\section{Spectro-temporal RF identification}

\label{sec:cnn}
\begin{figure}[t]
    \centering
    \subcaptionbox{Wi-Fi}{
        \includegraphics[width=.25\linewidth, height=1cm]{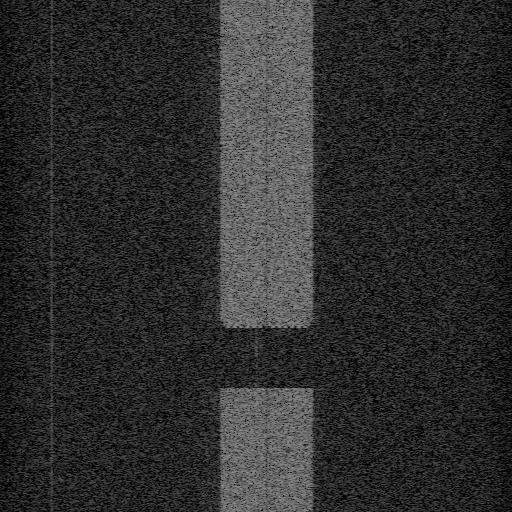}
    }
    \subcaptionbox{Bluetooth}{
        \includegraphics[width=.25\linewidth, height=1cm]{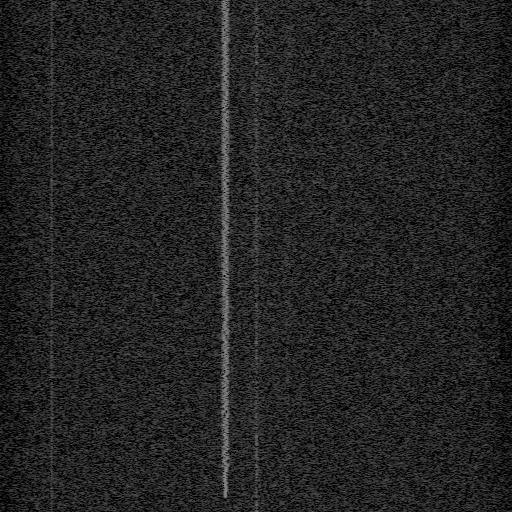}
    }
    \subcaptionbox{ZigBee}{
        \includegraphics[width=.25\linewidth, height=1cm]{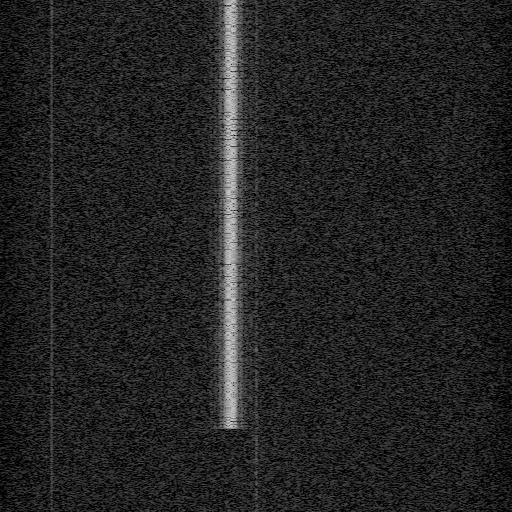}
    }
    \caption{Different emissions are distinguishable with computer vision-based representation of I\&Q data.}
    \label{fig:rep_example}
\end{figure}

\begin{figure}
    \centering
    \includegraphics[width=.9\linewidth, height=2.6cm]{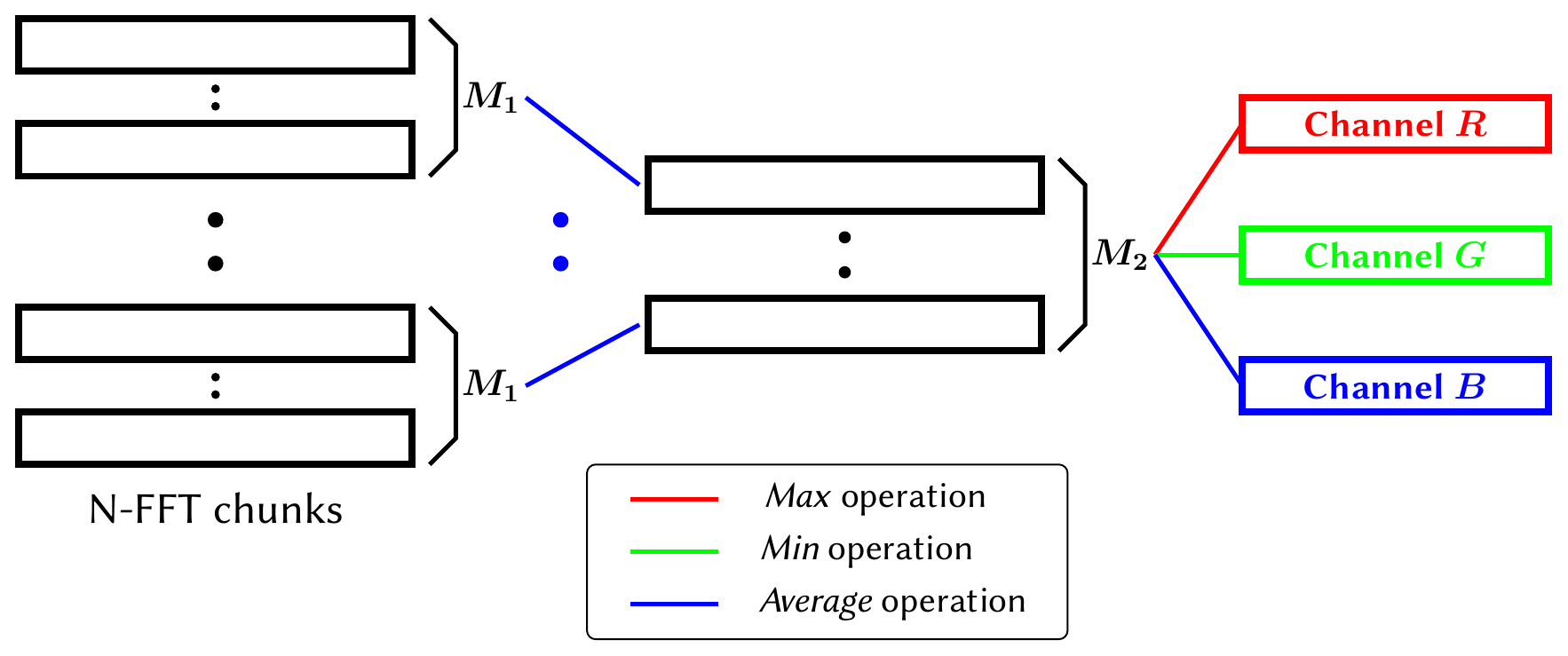}
    \caption{RF-centric compression mechanism. \textmd{Channels $R,G,B$ of a compressed RF image are mapped to outputs of \textit{Max, Min, and Average} operations.}}
    \label{fig:rf_compression}
\end{figure}

\begin{figure*}
    \centering
    \includegraphics[height=6.4cm]{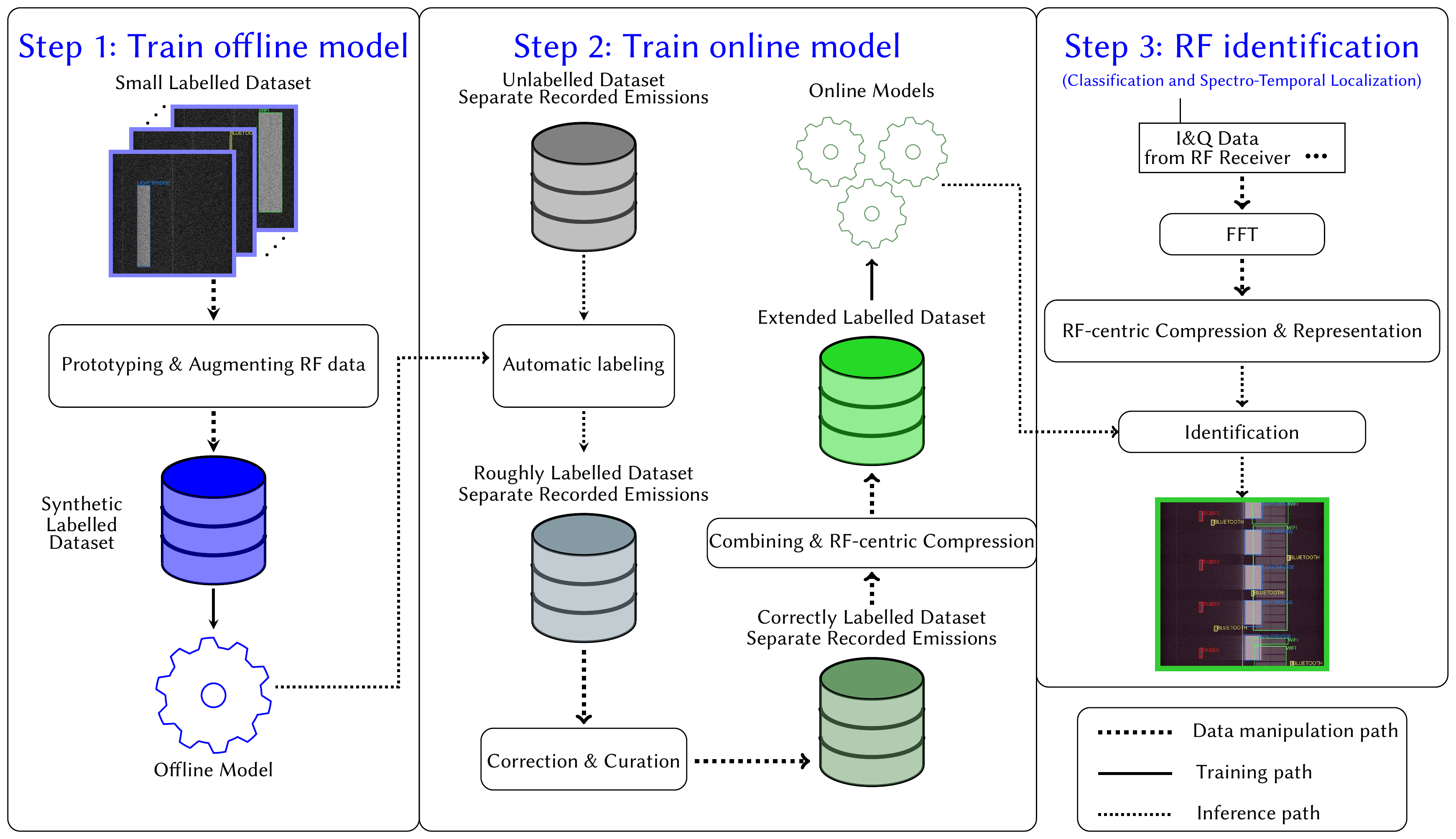}
    \caption{\system{}'s RF identification workflow.}
    \label{fig:detection_framework}
\end{figure*}

Our goal is to build a practical and scalable framework for RF identification. Towards this goal, we designed \system{} with the following specific objectives: accurate detection, spectro-temporal location, real-time processing and wideband spectrum support. To the best of our knowledge, previous work only focused on a subset of these objectives. We first provide
an overview of \system{}, highlighting the key challenges and our approach, followed with a more detailed description of the introduced mechanisms.

\subsection{Challenges and Elements of Approach}
\bfpara{Wideband spectro-temporal identification} Operation over a wideband spectrum is critical to most RF emissions identification applications, as most of today's wireless systems are flexible to operate and share fairly large bands (e.g., 2.4GHz ISM band is 80MHz wide and is home to various communications standards). Previous work has not fully addressed the problem of wideband processing for extracting accurate spectro-temporal information. 
For example, authors in~\cite{bitar17} attempted to observe the whole 2.4 GHz ISM band by using multiple classifiers that process samples in much narrower bands. This approach is hard to deploy in practice due to computation overhead and the complexity of synchronization between classifiers. Furthermore, the RF classifiers only infer the presence (detection) or the category (classification) of an emission in the samples, without revealing its spectro-temporal location. We tackle these problems by leveraging techniques from \textit{object detection} approaches of computer vision. We transform the wideband RF samples into a 2D time-frequency picture (\Cref{sec:input}) and run the RF-centric classifier to identify all individual and overlapping emissions by providing their categories as well as 2D positions (cf. an example of Wi-Fi packet detection in~\Cref{fig:detection}). 


\bfpara{Real-time processing} Despite that real-time processing is critical for practical RF recognition system, its analysis and resolution are still mostly lacking in the literature. We incorporate two features into the Deep Learning framework to solve this problem. In essence, we first designed a \textit{RF-centric compression} component (cf.~\Cref{sec:compression}) and applied it as the first layer of the learning network to reduce the amount of computation for the succeeding layers, while selectively preserving important features in the data. Second, we leveraged and \textit{enhanced} a state-of-the-art \textit{one-stage object detection} approach (details are discussed in \Cref{sec:onestage}) that processes the inference using a single network propagation that runs much faster in comparison to systems based on multiple complex pipelines. \ignore{To confirm the feasibility of the approach, we measured and analyzed \system{} developed using commercial NVIDIA GPU, and Software Defined Radio running on the popular Universal Software Radio Peripheral (USRP \cite{USRP}).}

\bfpara{Efficient data collection and training}
Any effective machine learning based identification framework requires a large amount of time to collect and train the dataset. This is specially true for deep learning approaches. In the RF domain, it is even more challenging to build a high-quality dataset of RF samples due to the huge amount of data that requires RF expert knowledge to be collected and annotated. For example, a 100 MHz wideband receiver can generate 800 Mbytes/sec\ignore{\footnote{Our receiver was running on a USRP X310~\cite{USRP} generating 8 bytes/sample.}}. As a result, obtaining a large, fully-labeled training dataset of high-speed RF data is particularly time-consuming and costly. To reduce the manual efforts for building the training dataset and extending it for future use, we use an iterative training process. First, we collected a small set of individual RF emissions and converted them to 2D images. We then applied various transformations on these images (e.g., shifting to different locations and adjusting the signal-to-noise ratio) to obtain a \emph{synthetic labelled dataset} and trained the first model, called \emph{offline model}. In the second step, we created a much larger set of RF samples by recording the over-the-air transmitted signals and employing the offline model to generate the appropriate annotations. The resulted dataset was expanded by additional RF combining operations (e.g., adjusting signal strength or generating signal collisions) to obtain an \emph{extended labelled dataset}. The \emph{online models} were then trained and evaluated leading to our final model for \system{}. The overall workflow of our system is depicted in~\Cref{fig:detection_framework}. It is emphasized that the RF-centric compression layer is only present in the online model. \ignore{In the following we discuss the core components of our system. We leave the description and evaluation of the training dataset for the offline model in~\Cref{sec:offline} and for the online model in~\Cref{sec:online}.}
\ignore{ On that account, \system{}'s final model is acquired using an iterative learning method: An \textit{offline model} is trained using efficiently generated data from prototyped and augmented RF emissions. Then, an \textit{online model} is trained on the large recorded over-the-air dataset labelled by the \textit{offline model} and fueled by careful amendments. The \textit{online model} is used as the final model of \system{} for practical RF identification in the wild.}

\subsection{Image-based RF Input Representation}
\label{sec:input}
Data representation is the first crucial step in designing a Deep Learning model. Recent work, such as RF classification~\cite{bitar17} or modulation recognition~\cite{vtechmodrec16, HCNNmodrec, CNNmodrec2}, directly fed I\&Q samples to the model in the form of a matrix of two rows consisting of in-phase and quadrature components.\ignore{ A naive approach would feed the I\&Q samples, which is represented as a matrix of ${2 \times N}$ where the rows include ${N}$ in-phase and quadrature samples, directly to the model. This method is leveraged by recent work in RFML modulation recognition \cite{XXX}.} While this method is simple, it is unclear\ignore{the proof of} how Deep Learning models can interpret this representation\ignore{ is lacking}. More importantly, it is challenging, in the short term, to improve and optimize the model due to incompatibility with most start-of-the-art Deep Learning algorithms specifically designed for other types of representation, particularly images, texts, and acoustics. 

Different wireless technologies are characterized by unique features such as frequency range, bandwidth, frame format, modulation and coding scheme.
For instance, Bluetooth (IEEE 802.15.1) uses channels of 1MHz with GFSK, DQPSK and DPSK modulations, while Wi-Fi (IEEE 801.11a/g/n/ac) leverages orthogonal frequency division multiplexing (OFDM) for 20--160 MHz channels consisting of multiple sub-carriers. Those features are recognizable by visual analysis of frequency and/or time domain of the recorded samples. 
Motivated by this observation, \system{} uses an image-based representation of RF data, 
enabling the use of numerous cutting-edge deep learning architectures specially designed for images~\cite{resnet16, VGG-2014, yolov1}. Specifically, we divide the stream of input RF samples into equal chunks and perform ${N}$-point Fast Fourier Transform (FFT) on each chunk. We group the FFT output of $M$ chunks to a $M\times{}N$ complex-valued matrix that represents \ignore{the view of} the spectrum in the time span of $M$ consecutive periods. An equivalent 2D grayscale image is then created by mapping between image pixels and matrix elements. In particular, if a matrix element on column $x$ and row $y$ has \ignore{a complex} value $m_{x,y}$, it is mapped to a pixel at coordinate $(x,y)$ with a value $p_{x,y}$ via a mapping function $f(z)$ as follows 
\begin{equation}
\small
p_{x,y}=f(A_{x,y}) := \gamma*(\min{(\max{(A_{x,y},A_{min})},A_{max})}-A_{min})
\label{eq:img-grayscaled}
\end{equation}
where $A_{x,y}=20*\log_{10}|{m_{x,y}|}-N_0$ representing the SNR of frequency bin $x$ of the $y$-th chunk measured in dB with respect to the noise floor $N_0$, and $\gamma=255/(A_{max}-A_{min})$ denoting the scaling factor for SNR-grayscale mapping.

\ignore{Examples of the above representation are depicted in~\Cref{fig:rep_example}, where grayscale images of IEEE 802.11 b/g (Wi-Fi), IEEE 802.15.1 (Bluetooth), and IEEE 802.15.4 (ZigBee) emissions were acquired.}
 Examples of IEEE 802.11 b/g (Wi-Fi), IEEE 802.15.1 (Blue-tooth), and IEEE 802.15.4 (ZigBee) emissions are depicted in~\Cref{fig:rep_example}.
We emphasize that while the phase information of $m_{x,y}$ is omitted and only the complex magnitudes $A_{x,y}$ are used for the grayscale image conversion, the  distinguishing RF features of different technologies such as bandwidth of emission textures are clearly visible.
We also note that the representation in~\Cref{eq:img-grayscaled} is only used for the \emph{offline model} (\Cref{sec:offline}), where all RGB color channels are assigned the same value (grayscale mapping). For the \emph{online model} (\Cref{sec:online}), each color channel is mapped to a different value determined by the \emph{RF-centric compression} (\Cref{sec:compression}), and the final system relies on the full RGB images.



\subsection{RF-centric Compression}
\label{sec:compression}
While the single network propagation with the \textit{one-stage object detection} can improve the detection speed, it is not enough for the real-time RF identification of wideband spectrum. We observed that our initial model trained for the RF emission dataset, took tens of milliseconds to process a 100 MHz incoming stream of RF samples that spanned only a few milliseconds. Increasing the duration of the input subsequently extends the spatial size of the neural network, thus making the detection slower. 

To circumvent the above issue, we designed an \textit{RF-centric compression} layer as the first layer of the \emph{online model}. This layer squeezes multiple RF input representations into a new RF representation that retains the important features of the original inputs. The working mechanism, illustrated in~\Cref{fig:rf_compression}, comprises two steps of compression. The first step involves compressing $M_1$ FFT output chunks into one average chunk, i.e., for every group of $M_1$ chunks of FFT output $\{m_{x,y_1}\}$, where $0\le{}x<N$ and $0\le{}y_1<M_1$, the layer computes the signal energy average $\frac{1}{M_1}\sum_{y_1}|m_{x,y_1}|^2$ on each individual frequency bin $x$ across the time dimension. 

\begin{figure}[t]
    \centering
    \subcaptionbox{\label{fig:comp_a}\footnotesize Wi-Fi}{
        \includegraphics[width=0.16\linewidth, height=1.5cm]{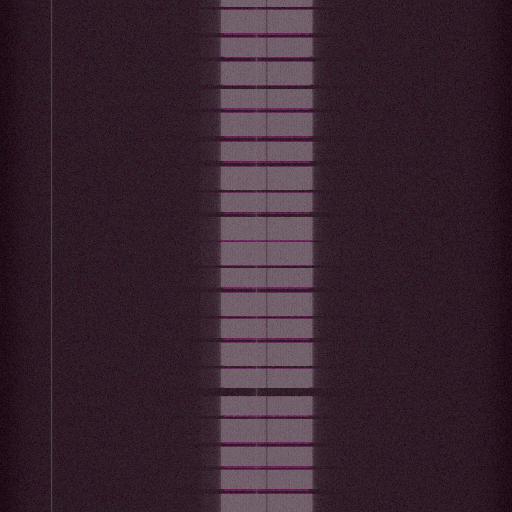}
    }
    \subcaptionbox{\footnotesize Bluetooth}{
        \includegraphics[width=0.16\linewidth, height=1.5cm]{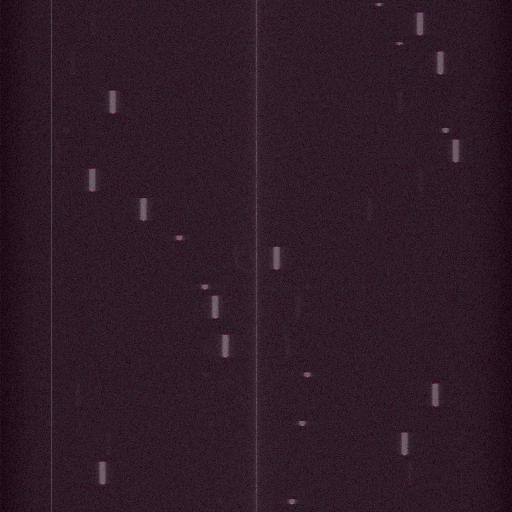}
    }
    \subcaptionbox{\footnotesize ZigBee}{
        \includegraphics[width=0.16\linewidth, height=1.5cm]{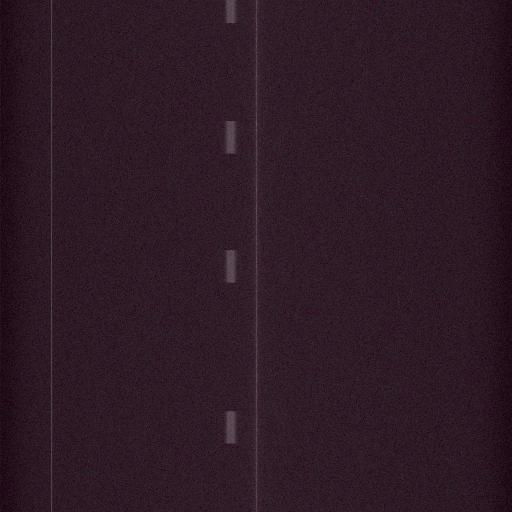}
    }
    \subcaptionbox{\footnotesize Lightbridge}{
        \includegraphics[width=0.19\linewidth,height=1.5cm]{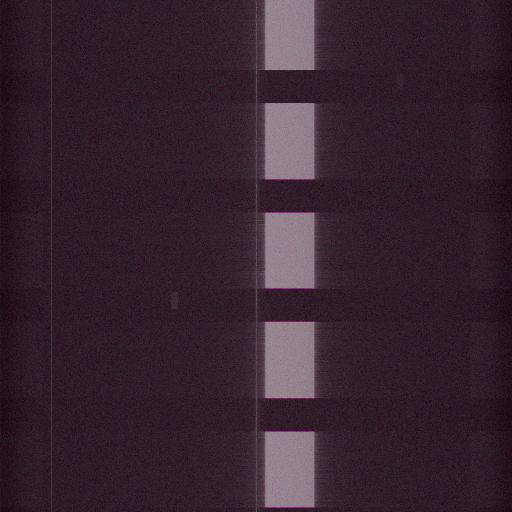}
    }
    \subcaptionbox{\footnotesize XPD}{
        \includegraphics[width=0.16\linewidth, height=1.5cm]{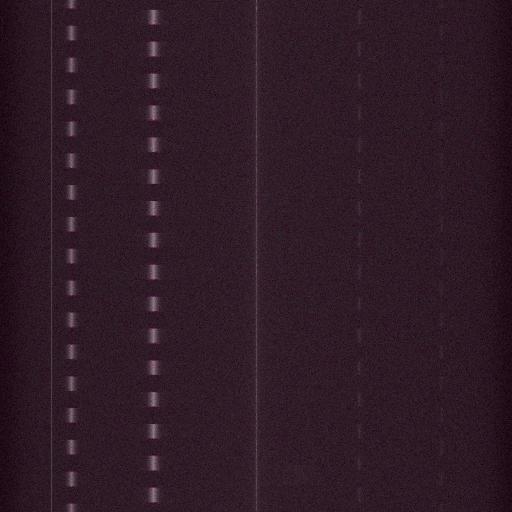}
    }
    \caption{Data resulting from RF-centric compression.}
    \label{fig:compress_example}
\end{figure}

Let $E_{x,y_2}$ denote the first step's results, where $y_2$ is the first step's output chunk index. Now in the second step, we again compress $M_2$ chunks of $\{E_{x,y_2}\}$, where $0\le{}y_2<M_2$, into one average chunk and obtain $E^{avg}_{x,y}=\frac{1}{M_2}\sum_{y_2}E_{x,y_2}$ for this second step's $y$-th output chunk. In addition to the average, we also compute the maximum and minimum value per frequency bin: $E^{max}_{x,y}=\max_{y_2}(E_{x,y_2})$ and $E^{min}_{x,y}=\min_{y_2}(E_{x,y_2})$. Each output chunk in the second step forms a row in the 2D picture, where a pixel is assigned an RGB color based on the following SNR to color channel mapping:
\begin{equation}
\begin{aligned}
R_{x,y} & = f(10\times \log_{10}E^{max}_{x,y} - N_0) \\
G_{x,y} & = f(10\times \log_{10}E^{min}_{x,y} - N_0) \\
B_{x,y} & = f(10\times \log_{10}E^{avg}_{x,y} - N_0)
\end{aligned}
\label{eq:img-colored}
\end{equation}
where $f(z)$ is the same mapping function used in \Cref{eq:img-grayscaled}.
Although some information is lost, the compression preserves important properties such as the high and low peaks in RF emanations that help distinguish \ignore{different} RF technologies, as well as crucial variations in signal strength in the three channels of the final representation. \Cref{fig:compress_example} shows clearly distinguishable ``compressed'' RF emissions. This compression layer is inspired by the computer vision's pooling method which filters the most important features in the data, thus relieving the computation load for the neural network. 

We also considered dropping I\&Q samples as an alternative approach for enhancing the real-time capability of the system. If enough samples are discarded, the processing rate can reach the incoming input data rate and the network can become real-time. However, it is highly possible that essential RF features for fingerprinting the emissions are neglected. Dropping too many samples will cause substantial degradation in the detection performance. Furthermore, very short RF transmissions (e.g., Wi-Fi ACK packet or ZigBee sensor data packet) can be frequently missed. Therefore, we contend that selective compression with the RF-centric first layer is a preferable choice.



\subsection{RF-centric Enhancements}
\label{sec:onestage}
Our Deep Learning framework is inspired by YOLO \cite{yolov1, yolov2, yolov3}, a \textit{one-stage object detection} method that has the most competitive detection speed in the current literature \cite{yolov4}. Unlike other \textit{two-stage detection} methods \cite{RCNN, fasterRCNN} that operate with slow, complex pipelines of region selection and object detection, YOLO unifies those stages into a single forward propagation by a convolutional neural network architecture. Towards achieving real-time RF identification, we also show that YOLO can be optimized to gain significant improvements in speed and detection precision, based on the observable and distinctive characteristics of RF emissions.

While some prior-work considered one-stage objects detection for RF identification~\cite{oshea_yl17, Lo_2020_WACV}, it nonetheless lacks the mechanisms for achieving real-time and wideband performance, and does not address the issues of detecting  multiple RF technologies. Furthermore, data collection methods for a large RF training dataset are absent. In this section, we describe our enhancements to YOLO to address the aforementioned problems and achieve our goal.

\bfpara{RF-centric Anchor Boxes}
The YOLO neural network targets to output a set of \textit{bounding boxes}, each for a single object in the input image. All features including emissions and noise in every time and frequency slot need to be considered for detection. On that account, the network divides the input into ${S \times S}$ grid cells, each generating ${B}$ bounding boxes predicting the objects whose centers are located within that cell. To produce the prediction, YOLO uses a set of anchor boxes for each grid cell, which are the pre-defined bounding boxes with specific sizes, as references for the predictions of objects. They are the fundamental components that enable the capabilities to capture objects of different aspect ratios in many state-of-the-art object detectors \cite{yolov3, yolov4, fasterRCNN}. Therefore, it is important for the anchor boxes to be well-suited for the objects that a model learns and predicts. We observed that visualized RF emissions typically have highly-varying sizes, instead of fixed sizes as for real-life objects. Hence, using \textit{RF-centric} anchor boxes can enhance the learning process as well as generate more precise detection. For that reason, we replaced the default image-based anchor boxes used in YOLO (acquired from the ImageNet dataset \cite{imagenet_cvpr09}) with our \textit{RF-centric} anchor boxes generated by $K$-means clustering on the training dataset as in \cite{yolov2}. As we discuss later, \textit{RF-centric} anchor boxes can boost YOLO's performance in the extreme cases that demand significantly precise detection.

To tell whether a bounding box associates with an RF emission, a confidence score is computed for each box, which is a multiplication of the boolean indicator for the predicted emission presence (${P(\mathrm{obj})\in\{0,1\}}$) and the Intersection-over-Union (IoU)\footnote{Also called Jaccard Index, Intersection-over-Union measures the similarity between a predicted and a ground-truth bounding box by taking the ratio of the overlapping area over the merging area of the boxes.} of that box and the ground truth. When there is an existing emission in a cell, the confidence score is equal to the IoU, otherwise it is zero. A bounding box is predicted with the conditional probabilities (${P(\mathrm{RFclass}_l|\mathrm{obj}), l \in[1\ldots{}C]}$) for all ${C}$ different RF technologies (classes). The position of the box is described by four parameters:  the center coordinates of the box $(x_c, y_c)$ relative to the cell position, and the width and height ${w, h}$ relative to the size of input image.


\begin{figure}
    \centering
    \includegraphics[width=\linewidth, height=3.8cm]{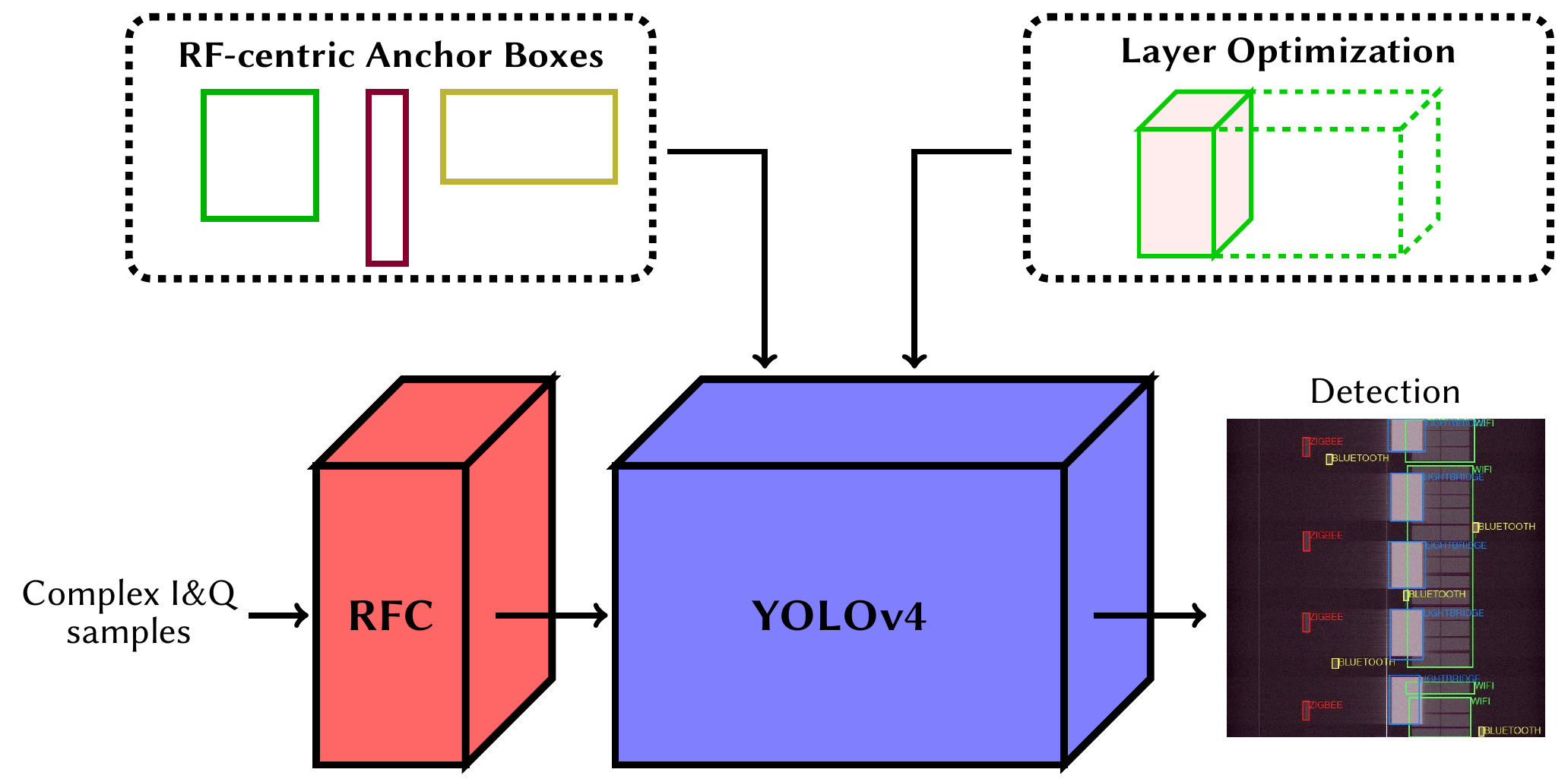}
    \caption{\smodelnew{} model consists of the RF-centric compression layer (RFC) and the optimized version of YOLOv4 \cite{yolov4}.}
    \label{fig:syl_4}
\end{figure}

\bfpara{Layers Optimization} 
YOLO uses a deep convolutional neural network that incorporates three detection layers that divide the input data into grid cells of three different scales. Predictions at larger scales are made possible by combining upsampled feature maps of previous predictions with finer-grained feature maps from the earlier layers in the network. This allows better predictions for RF emissions which are significantly different in size in the RF representation, e.g., narrow, short bluetooth vs. long and wide Wi-Fi RF emanations. Although YOLO can perform object classification at high frame rate for a generic computer vision task, it remains challenging to achieve real-time operation for \system{} with the off-the-shelf complex YOLO architecture. By adding the RF-centric compression layer with a large compression factor, we speed up the system at the expense of information loss. To achieve high accuracy while enabling real-time processing, we optimized YOLO's convolutional layers by selectively reducing the volume of convolutional filters based on an observation that visualized RF emissions are sharp and simpler than real-life objects (which are the initial targets for YOLO design). In other words, the visualized RF emissions have less features that can be extracted, leading to a smaller volume of convolution filters sufficient for detection. We reduced the filter volume stage-by-stage until reaching significant increase of the validation error:
\begin{equation}
U_i = U_{i-1}\times (1-\sigma^i)
\label{eq:filter-adjust}
\end{equation}
where $\sigma=0.5$ and $U_i$ is the filter volume at stage $i$. We stopped decreasing the layer volume after $i=2$, which resulted in the total reduction of 62.5\%. This modification preserves the detection performance, while speeding up the inference by more than $2.2 \times$, allowing lower compression factor for better performance.

During the training process, YOLO optimizes a loss function comprising three elements to penalize the errors in the box coordinates, output confidence score, and class probabilities. The mean squared error loss is used for box coordinates, while the binary cross-entropy loss is used for the others. It should be noted that the total loss to be optimized is the sum of the losses computed at the three detection layers. Whereas in the prediction process, each of the three detection layers will output a 3-D matrix (or \textit{tensor}) of size ${S_i \times S_i \times [B \times (1 + 4 + C)]}$ where ${S_i \times S_i}$ is the grid size of the ${i}$-th scale. There are often cases when large RF emissions (e.g., Wi-Fi) spanning multiple cells can result in numerous predicted boxes. We used non-maximal suppression \cite{yolov1} to remove redundant boxes which have IoU with the main predicted box (i.e., one has the highest confidence score) exceeding ${0.5}$ if they share the same predicted class.

The latest YOLO version v4 has achieved significant improvements for object detection~\cite{yolov4}. This method incorporates various recent Deep Learning techniques that improve the training process (\ignore{of which they called "Bag of Freebies", }such as data augmentation) and the inference process (\ignore{of which they called "Bag of Specials", }such as attention model and post-processing methods)\ignore{ with the previous version}. Those improvements boost the detection accuracy, with a modest reduction in inference speed. \ignore{Later, we also investigate}We apply our RF-centric enhancements on YOLOv4 in order to finalize the foremost detection \ignore{model for the}\textit{online model} as depicted in \Cref{fig:syl_4}.

\begin{figure}[t]
    \centering
    \subcaptionbox{\label{fig:syn_a}Spectro-temporal moving}{
        \includegraphics[width=.14\linewidth]{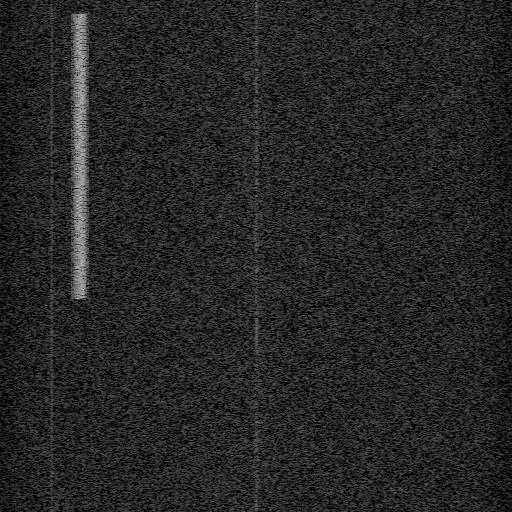}
        \includegraphics[width=.14\linewidth]{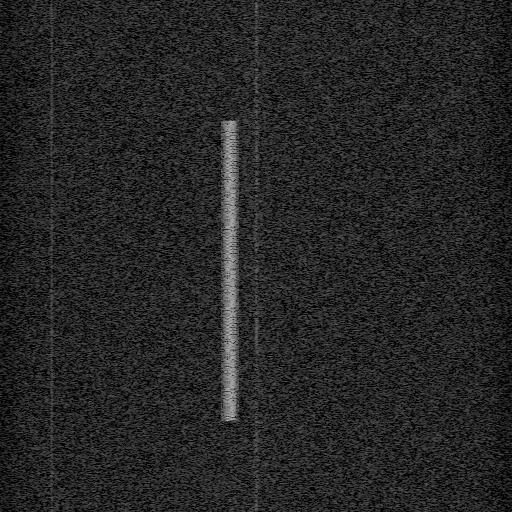}
        \includegraphics[width=.14\linewidth]{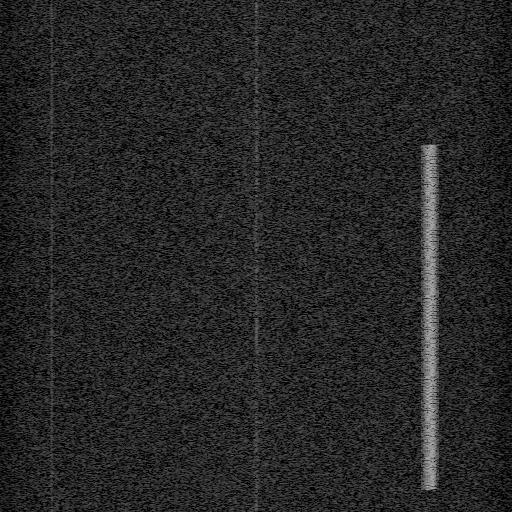}
    }
    \subcaptionbox{\label{fig:syn_b}Altering emission length}{
        \includegraphics[width=.14\linewidth]{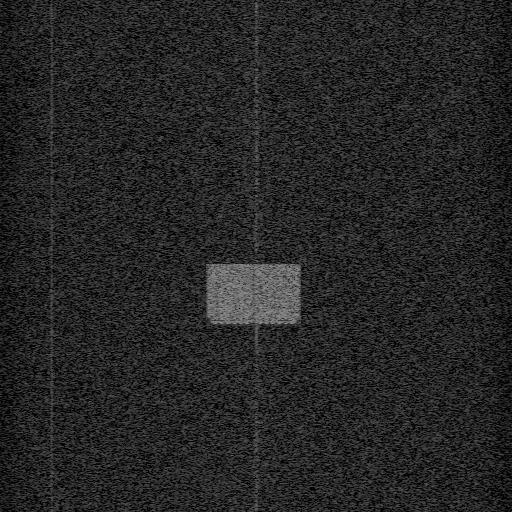}
        \includegraphics[width=.14\linewidth]{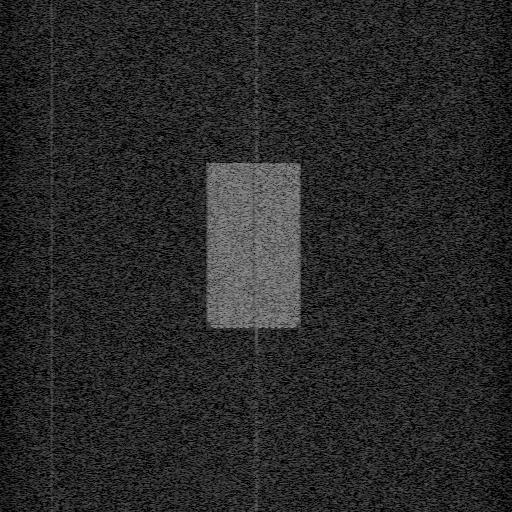}
        \includegraphics[width=.14\linewidth]{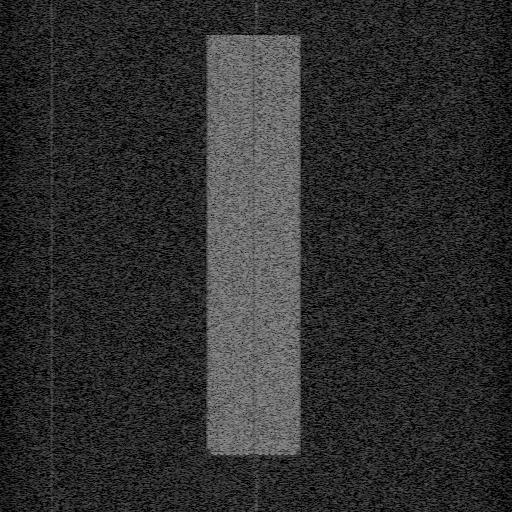}
    }
    \subcaptionbox{\label{fig:syn_c}Varying emission SNR}{
        \includegraphics[width=.14\linewidth]{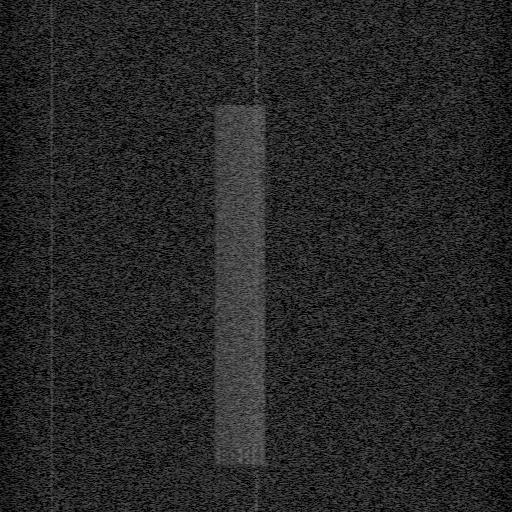}
        \includegraphics[width=.14\linewidth]{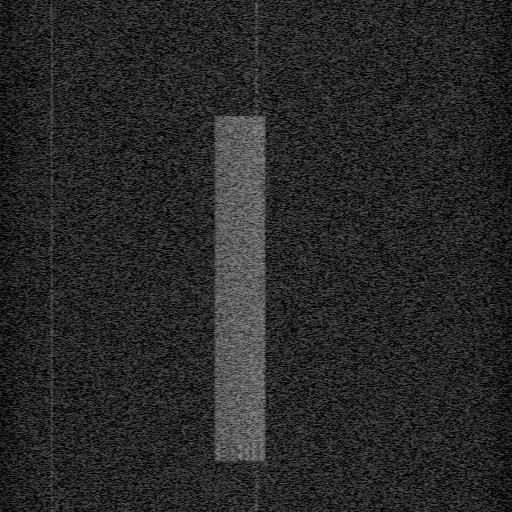}
        \includegraphics[width=.14\linewidth]{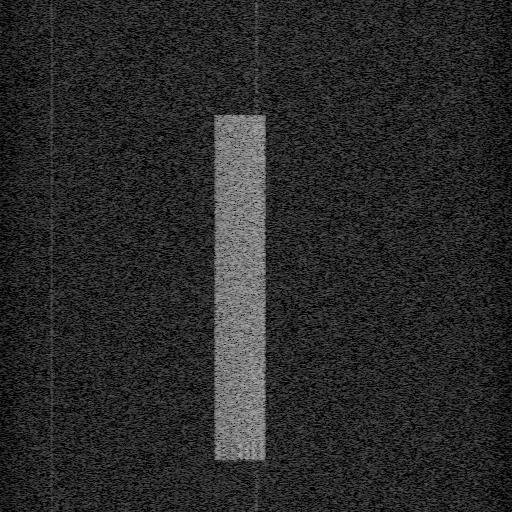}
    }
    \subcaptionbox{\label{fig:syn_d}Simulating RF collisions}{
        \includegraphics[width=.14\linewidth]{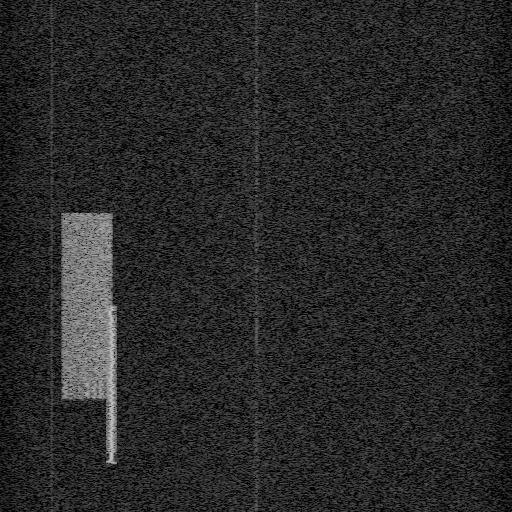}
        \includegraphics[width=.14\linewidth]{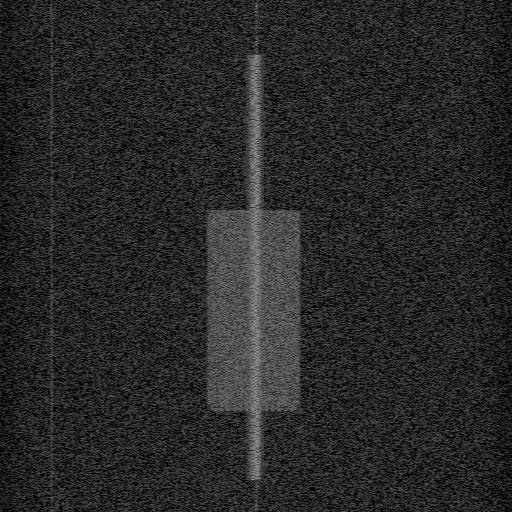}
        \includegraphics[width=.14\linewidth]{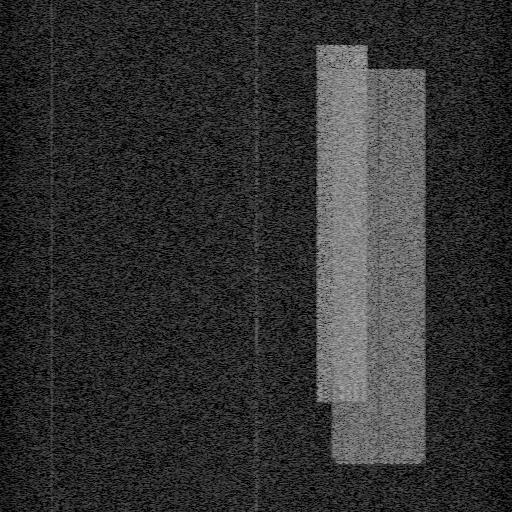}
    }
    \caption{Examples of augmented RF images generated based on prototypes of RF emissions for offline model.\ignore{ \textmd{With prototypes from recorded RF data, we have a full control over the \textbf{(a)} position, \textbf{(b)} length, \textbf{(c)} brightness, and \textbf{(d)} overlapping pattern of the objects representing the emissions.}}}
    \label{fig:syn_example}
\end{figure}

%% file: synthetic.tex
\section{Offline model}
\label{sec:offline}
In this section, we present the process of building the \textit{offline model}, an important step towards achieving adequate curated, labelled RF data and train practical deep learning models. We show that by using a small amount of labelled data and efficient data manipulation and augmentation techniques, we can achieve a synthetic training dataset that is sufficient for transfer learning to support automatic labelling of a new much larger dataset. We note that since the offline model is only used internally in \system{} to assist the online training, the real-time requirement is relaxed and the RF-centric compression layer is disabled.

\begin{figure*}
    \centering
    \includegraphics[width=\textwidth,height=5.5cm]{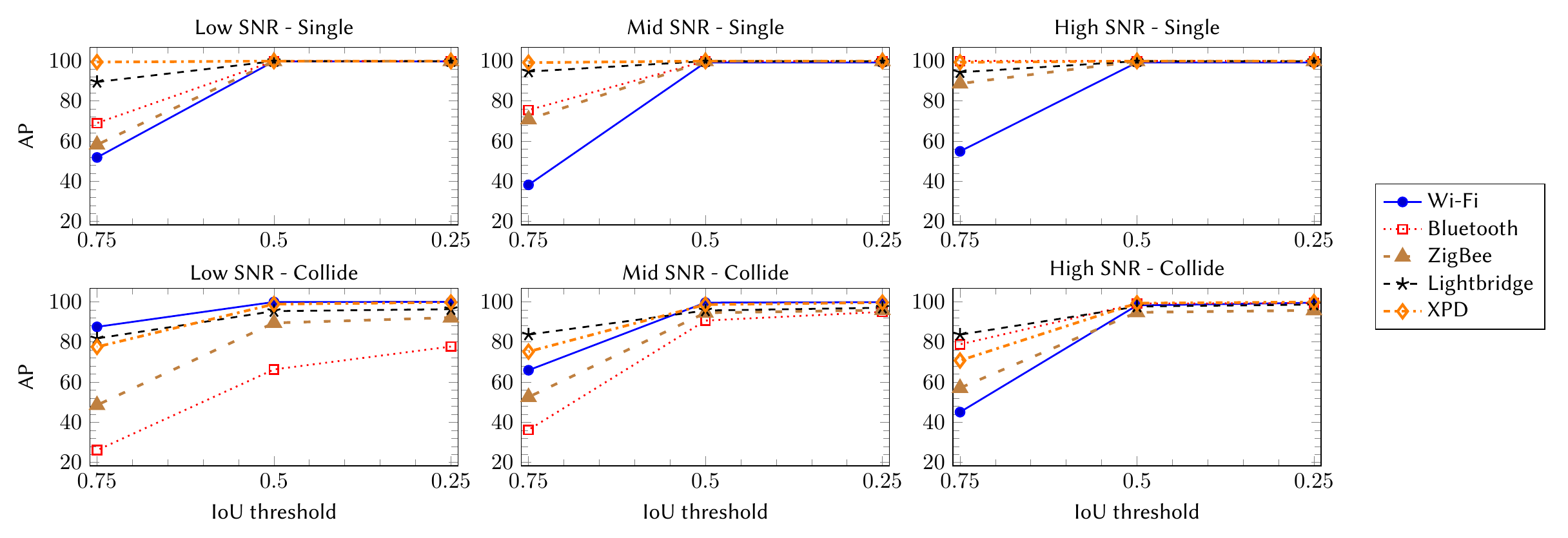}
    \caption{Test results for the \textit{offline model} with regards to RF classes, SNRs, and IoU thresholds.}
    \label{fig:synthetic_result_class}
\end{figure*}

\subsection{Synthetic RF dataset} 
\ignore{Authors in \cite{NguyenVVN2019} utilized the artificially generated RF transmissions for deep transfer learning in wireless collisions detection. Inspired by that, we made some improvements to efficiently achieve a more realistic dataset.} We developed a technique for generating synthetic labelled samples to bootstrap the transfer learning process. 
We collected a small dataset of a few labelled RF images, then created \textit{prototypes of emissions} cropped from the images. Based on those prototypes, we generate new RF images by applying several image transformations: (1) adjusting the object's brightness to alter the transmission SNR, (2) changing the length of object by cropping or concatenating to vary the transmitted duration, and (3) moving the object to different locations within the image to simulate various spectro-temporal positions. These image-based transformations allow us to mimic the real RF data and efficiently generate sufficient training samples as depicted in \Cref{fig:syn_example}. It is worth noting that while performing those transformations, we systematically generate all the emission annotations, without manual effort, which include the RF category and the bounding box's four positioning parameters.

We created a dataset of 99,330 RF images of size $512 \times 512$, where each image captured the view of a 100 MHz spectrum over $2.62\mathrm{ms}$ time span with resolution of $N=512$ frequency bin and $M=512$ time slots. There are 47,830 samples of single RF emission and 51,500 samples of overlapping emissions in the synthetic dataset. The whole dataset contains 150,830 fully labeled synthetic RF emissions of five wireless radio technologies: IEEE 801.11 (Wi-Fi), IEEE 802.15.1 (Bluetooth), IEEE 802.15.4 (ZigBee), Lightbridge (DJI protocol for robust aerial communications~\cite{DJI-lightbridge}), and XPD (Samson's cordless microphone systems~\cite{samson-wmic}). 
In addition to single RF emission samples, we also synthesized collisions (examples in \Cref{fig:syn_d}) which are common in real scenarios.  
The dataset is split into training set, validation set, and test set with the ratio $0.64:0.16:0.2$ correspondingly.  

We reused the open-source YOLO implementation\footnote{https://github.com/AlexeyAB/darknet} to train our offline deep learning model. In this implementation, Batch Normalization~\cite{batchnorm} was exploited to significantly improve the convergence and remove the requirements for regularization. Inputs were scaled to $608 \times 608$ before passed into the network. We used a batch size of 64 and learning rate $\alpha=\num{1e-3}$. We utilized Stochastic Gradient Descent (SGD) optimizer with momentum $\beta = 0.9$ and weight decay $\lambda = \num{5e-4}$. The neural network was trained on a NVIDIA GeForce GTX 1080 GPU with the first 53 convolutional layers pretrained on the ImageNet~\cite{imagenet_cvpr09} dataset to utilize the visual feature maps learned from various real life objects.

\begin{table}
\small
  \caption{Results of the \textit{offline model} on synthetic data.}
  \label{tab:synthetic_result}
  \begin{tabularx}{\linewidth}{m{3cm}|m{1.4cm}m{1.4cm}m{1.4cm}}
      Test set & \thead{mAP\\ $IoU_{0.25}$} & \thead{mAP\\ $IoU_{0.5}$} & \thead{mAP\\ $IoU_{0.75}$}\\ \hline
    Single emission & \thead{ 99.14} & \thead{ 99.13} & \thead{ 74.37}\\
    Colliding emissions & \thead{ 90.69} & \thead{ 88.28} & \thead{ 65.13}\\
    Total & \thead{ 91.72} & \thead{ 89.67} & \thead{ 66.42}\\
\end{tabularx}
\end{table}

\subsection{Evaluation} Existing work in RF classification uses \textit{class accuracy} as the main evaluation metric. However, this metric is not capable to evaluate spectro-temporal \textit{localization}, and also not robust against unbalanced classes. In this work, we use \textit{mean Average Precision} (mAP, the mean of the Average Precisions of all classes), a popular metric for object detection~\cite{pascal-voc}. mAP is calculated with three different Intersection-over-Union (IoU) thresholds (0.25, 0.5, 0.75 -- denoted by $IoU_{0.25}$, $IoU_{0.5}$, $IoU_{0.75}$). A transmission is detected if the IoU with the ground truth exceeds the threshold.

\Cref{tab:synthetic_result} shows the overall test result, as well as results on the two separate sets of single (non-overlapping) emissions, and emissions with collisions (overlapping). We can see that the \textit{offline model} has higher than 90 mAP with IoU threshold of 0.25 in all cases. Especially, for single emissions, it achieves very high mAP of 99.14 and 99.13 for IoU threshold of 0.25 and 0.5, respectively. In the case of collisions, the mAP degrades to 90.69 for $IoU_{0.25}$ and 88.28 for $IoU_{0.5}$. For IoU threshold of 0.75 which requires stricter localization, the mAP decreases to 74.37 and 65.13 for single emissions set and collisions set respectively.

\begin{figure}
    \centering
        \includegraphics[width=.12\linewidth,height=1.4cm]{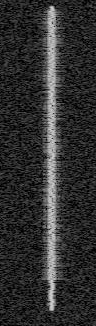}
        \hspace{0.6cm}
        \includegraphics[width=.12\linewidth,height=1.4cm]{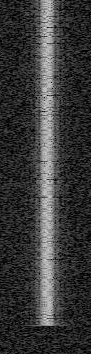}
    \caption{Example of a confusion that causes misclassification when using the \textit{offline model} for real RF emissions. \textmd{A saturated Bluetooth emission (Left) that emits leakage can be confused with a Zigbee emission (Right), and vice versa.}}
    \label{fig:syn_error}
\end{figure}

\Cref{fig:synthetic_result_class} shows how the performance varies with SNR for different RF technologies. Emission SNRs are categorized into three classes: Low (5-14 dB), Mid (15-24 dB), and High (25 dB and above). It can be seen that for $IoU_{0.5}$ and $IoU_{0.25}$, the Average Precisions (APs) for single RF emissions are higher than 99, sharing the same patterns with the overall results. For $IoU_{0.75}$, the APs decrease substantially (except for some High-SNR classes which still maintain good result). Interestingly, we can see that most of the classes have higher AP with $IoU_{0.75}$ as the SNR increases. The exception is Wi-Fi, in which we observed that the detection boxes for some particular long emissions that almost cover the height of a picture (as Wi-Fi emissions are 20 MHz wide, they cover approximately 20\% picture) are remarkably smaller than the ground truth, and thus count as false positives for $IoU_{0.75}$ threshold. We argued that is a drawback of YOLO when trained with non-RF-centric anchor boxes and addressed this when training the \textit{online model} with the optimized version of YOLO which has an improved performance. 

In general, the results for the collisions set are worse than the other set. Nonetheless, most of the classes still have higher than 80 AP with $IoU_{0.5}$ and $IoU_{0.25}$ (except Low-SNR Bluetooth, which are small and easily overshadowed when colliding with other stronger, much larger RF emissions). Also, it is evident from \Cref{fig:synthetic_result_class} that classes with larger size in RF representation (Wi-Fi, Lightbridge) tend to have higher and more stable AP across the three IoU thresholds (Wi-Fi is again an exception with $IoU_{0.75}$ due to the limitations of the detection for significantly long objects) because it is easier to recognize the larger object when two objects with different sizes overlap.

%% file: real-time.tex
\section{Online model for Real-time system}
\label{sec:online}
In this section, we describe the development of \system{}, built upon the \textit{online model}, which was trained on the extended dataset of recorded RF emissions. In order to build the training dataset, we bootstrapped the \textit{offline model} to automate the sample annotation. Additionally, we show that training data can be scaled efficiently by combining different RF recordings together. The final system achieves real-time operation on wideband RF data, thanks to the formerly described \textit{One-stage Object Detection} integrated with \textit{RF-Centric Compression} and \textit{RF-centric} model enhancements. \Cref{tab:comparison} compares \system{} with prior work, indicating that our system is the first deep-learning based approach capable of real-time, wideband spectro-temporal identification of real RF emissions. Through the development and evaluation on existing commercial hardware, we show that \system{} is a practical and cost-effective system.

\begin{table}
  \caption{2.4 GHz wireless emitters used in this work.}
  \label{tab:devices}
  \begin{tabularx}{\linewidth}{m{1.6cm}m{3.4cm}m{3.5cm}}
    \hline
    Technology & Device & Frequency range\\
    \hline
    Wi-Fi & \small{TP-LINK TL-WN722N} & \small{2.412 - 2.462 GHz}\\
    Bluetooth & Avantree DG60 & \small{2.402 - 2.480 GHz} \\
    ZigBee & TI CC2430EM & \small{2.400 - 2.483 GHz}\\
    Lightbridge & DJI Phantom 4 & \small{2.404 - 2.470 GHz}\\
    XPD & \small{Samson XPD2 Lavalier} & \small{2.404 - 2.476 GHz}\\
    \hline
\end{tabularx}
\end{table}

\subsection{Online Model}
\label{sec:train_eval_onl}
In this section, we evaluate and compare the performance of a minimally modified YOLO retrained on RF data (\textit{rf}YOLO) and our optimized models to determine the best option for the \textit{online model}. Comparing to a minimal augmented/ retrained YOLO, our optimized models have two advantages: Faster inference and are more RF-centric. The ability to quickly generate predictions allows the system to either handle more data (i.e., by increasing the sample rate which can cover a larger bandwidth) or preserve more features to predict more precisely (i.e., by reducing the compression factor of the RF-centric compression layer). For this purpose, we removed several convolutional filters, due to the fact that RF emissions are coarse and have significantly less features than real-life objects. Furthermore, the optimized models are more RF-centric with the anchor boxes derived (using k-means clustering) from the RF dataset instead of computer vision-based dataset of real-life objects (e.g., ImageNet \cite{imagenet_cvpr09}), as in YOLO models. Using the anchor boxes that better reflect the shape variations of RF emissions would help to provide more precision in the detection. We applied the modifications to both YOLOv3 and YOLOv4, and retrained on RF data to generate two optimized models: \smodelold{} and \smodelnew{}, respectively. The impact of optimizations is analyzed and discussed below, through comparison between the models. 

The quality of training datasets is the key factor for a good deep learning model. Although a model learned from synthetic RF data can learn certain underlying RF features, it is not capable to capture some specific RF variations from over-the-air wireless emissions, as well as to recover from incorrect assumptions in the synthetic data. When we tested the \textit{offline model} on several recorded RF emissions, the common error was misclassification caused by confusions resulting from unconsidered RF factors in the synthetic data such as out-of-band leakage (\Cref{fig:syn_error}). Fortunately, that type of error often introduces some patterns  which can be quickly corrected with automated tools. To that account, the \textit{offline model} can be bootstrapped for automatic labeling to achieve a large dataset of recorded over-the-air RF emissions. 


\begin{figure}
    \centering
        \includegraphics[width=.7\linewidth,height=3cm]{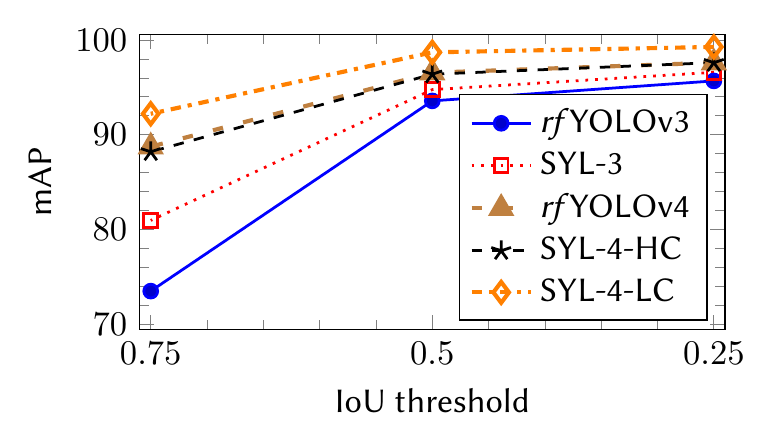}
    \caption{Results of \smodelold{, \smodelnew{}} compared with the original models. \textmd{\smodelnew{}-HC, \smodelold{} and YOLO use higher compression parameters $M_1=M_2=5$ to allow the real-time capabilities of \system{} \ignore{using YOLO models}. \smodelnew{}-LC uses lower compression with $M_1=3,M_2=4$ to exploit the faster inference of \smodelnew{}.}}
    \label{fig:real_compare}
\end{figure}

The training dataset for the online model was built as follows. We collected RF emissions from five radio technologies: Wi-Fi, Bluetooth, ZigBee, Lightbridge and XPD.  First, for our goal of obtaining clean recordings with minimal interference in the 2.4 GHz ISM band, we recorded different RF emission types separately. 
Using the \textit{offline model} for automatic labeling of the dataset, we observed that 41 percent of RF images needed some manual corrections and box adjustments. This result, on one hand, shows that our iterative training approach saved us substantial amount of time and  effort building a large curated and labelled dataset. On the other hand, we see that it is difficult for the \emph{offline model}, which is trained based on solely synthetic data, to achieve a high-quality detection.

A good deep learning model often requires considerable amount of training data to avoid overfitting and to generalize well. In the second step, we adjusted the transmission SNR in three ranges (measured in dB units): Low (5-14), Mid (15-24), and High (25 and above). In addition, we combined the separate recordings' I\&Q samples in the time domain to achieve a much larger dataset with coexisting and overlapping patterns of different types of RF emissions without incurring additional effort in labeling, by re-using the corrected and curated annotations from the first step. We note that in contrast to the synthetic dataset used for the offline model, the extended dataset for the online model was collected and synthesized using RF-based manipulations. 
As the final step before training the \emph{online model}, we enabled the RF-centric compression layer with parameters $M_1=M_2=5$ to produce a dataset of 253,397 compressed RF images of size $512 \times 512$. It is noted that the compression parameters (integer-valued) should be roughly equal to balance the visual effects of the two levels of compression. Additionally, The total compression factor needs to be sufficient to extend the duration of a data sample to meet the detection time. We split the dataset with ratio $0.64: 0.16 : 0.2$ for the training, validation, and test sets, respectively. The same training hyperparameters were used as for the \textit{offline model}. We emphasize that the \emph{online model} was trained completely from scratch, in contrast to the \emph{offline model} which was trained by transfer learning with pre-trained preceding layers.

\begin{table}
\small
  \caption{Detection time of the models for an input instance.}
  \label{tab:det_time}
  \begin{tabularx}{\linewidth}{lcccc}
  \hline
    Model & \textit{rf}YOLOv3 & \smodelold{} & \textit{rf}YOLOv4 & \smodelnew{} \\
    Detection time & 44.19 ms & 17.23 ms & 51.35 ms & 22.96 ms \\
  \hline
\end{tabularx}
\end{table}


\Cref{fig:real_compare} depicts the performance of the optimized and original models on the test set. It is evident that \smodelold{} achieves a substantial improvement of more than 6 mAP with $IoU_{0.75}$, and approximately 1 mAP with $IoU_{0.25}$ and $IoU_{0.5}$, compared to \textit{rf}YOLOv3, thanks to the use of RF-centric anchor boxes. We can also observe that \textit{rf}YOLOv4 outperforms \textit{rf}YOLOv3 with 15 mAP higher for $IoU_{0.75}$ and 2 mAP higher for the other thresholds. Meanwhile, our \smodelnew{} with higher compression (\smodelnew{}-HC in \Cref{fig:real_compare}) has comparable precision with \textit{rf}YOLOv4, both having higher than 88 mAP, 96 mAP, and 97 mAP for $IoU_{0.75}$, $IoU_{0.5}$, and $IoU_{0.25}$, respectively. More importantly, besides having competitive performance, \smodelold{} and \smodelnew{} models are more than $2.2 \times$ faster than the corresponding original models, as shown in \Cref{tab:det_time}. Consequently, we chose to re-train \smodelnew{} as the final \textit{online model} with lower compression factors (\smodelnew{}-LC in \Cref{fig:real_compare}) for further detection improvements.

We adjusted the compression factors to $M_1=3,M_2=4$ to generate the final dataset, which comprises 528,758 compressed RF images of size $512 \times 512$. After that, the dataset was split and used to train the final \textit{online model} by a similar process as with the previous dataset. Using the final model, we achieved 99.27, 98.70, and 92.21 mAP for $IoU_{0.25}$, $IoU_{0.5}$, and $IoU_{0.75}$, respectively. Most importantly, \smodelnew{}-LC got a considerable improvement of more than 3 mAP for the most difficult case $IoU_{0.75}$, compared to when trained with higher compression parameters, as depicted in \Cref{fig:real_compare}.

\begin{figure}
    \centering
        \includegraphics[width=.9\linewidth,height=1.5cm]{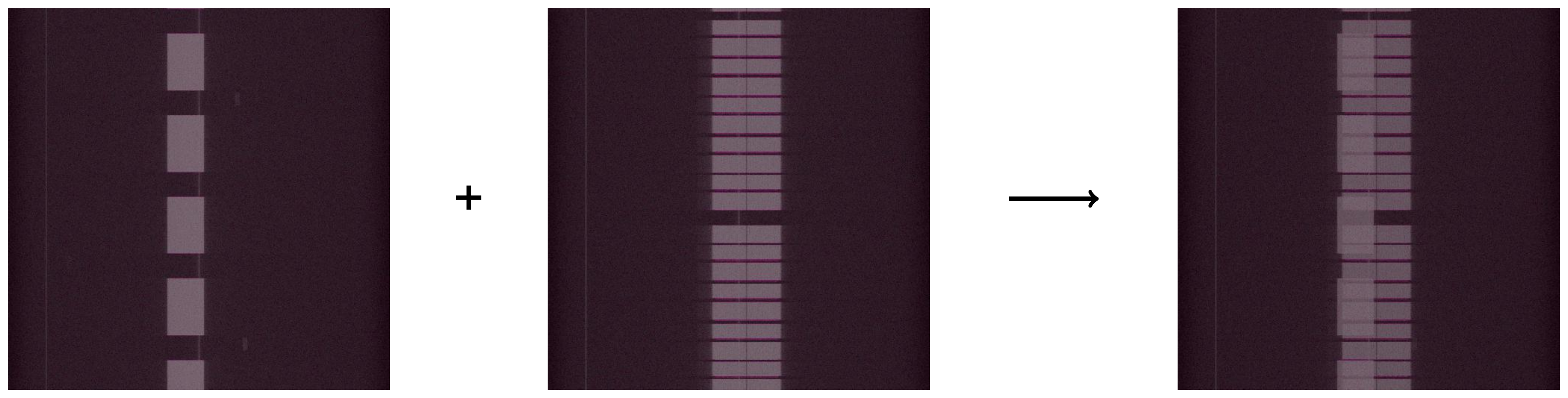}
    \caption{Combining RF recordings for new pattern. \textmd{I\&Q samples are added together in time domain, assuming \ignore{recording} in the Additive White Gaussian Noise (AWGN) environment.}}
    \label{fig:real_combining}
\end{figure}

\begin{figure*}
    \centering
    \includegraphics[width=\textwidth,height=3cm]{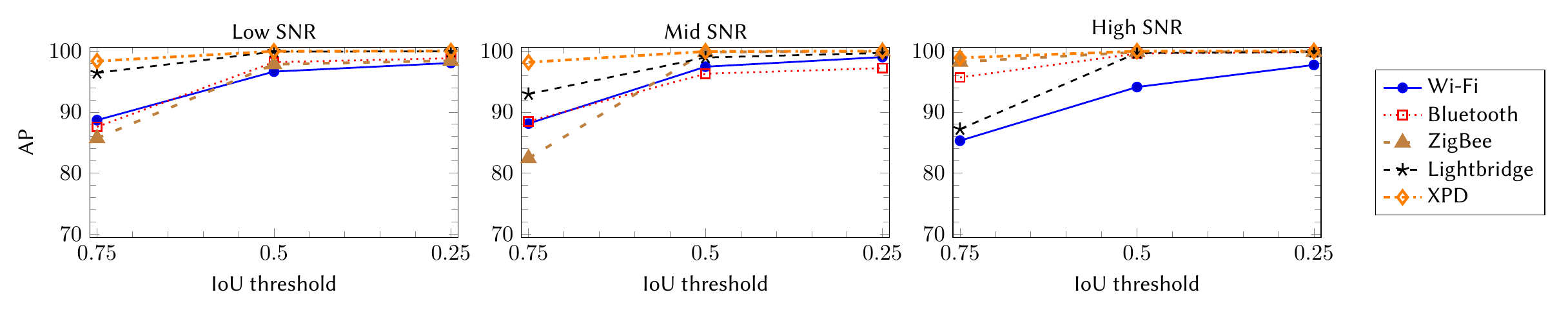}
    \caption{Test results for the \textit{online model} with regards to RF classes, SNRs, and IoU thresholds.}
    \label{fig:recorded_result_class}
\end{figure*}

\Cref{fig:recorded_result_class} provides more details of the results. It is evident that XPD is the most recognizable category whose AP maintains above 98 regardless of \ignore{variations of }IoU threshold and SNR. In addition, all the classes have higher than 80 AP for $IoU_{0.75}$, and higher than 90 AP for $IoU_{0.25}$ and $IoU_{0.5}$, across different SNRs. There is no significant difference between the results for low and middle SNRs, whereas Bluetooth and ZigBee gain substantial increases of more than 7 mAP for high SNR. \ignore{We also evaluated the false alarm probability of the model, where an emission is sensed correctly if the IoU of the predicted bounding box and the ground truth exceeds a threshold. With this metric, our model achieved desirable values of 4.4 \%, 5.1 \% and 9.8\% for  $IoU_{0.25}$, $IoU_{0.5}$ and $IoU_{0.75}$, respectively.}

\begin{figure}
  \begin{minipage}{0.45\linewidth}
     \centering
     \includegraphics[width=1.5in, height=.8in]{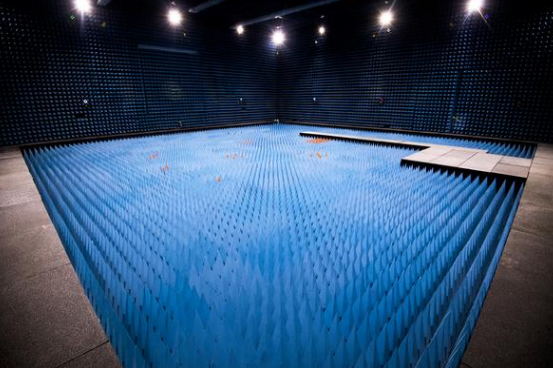}
    \caption{Anechoic chamber.}
    \label{fig:chamber}
  \end{minipage}\hfill
  \begin{minipage}{0.45\linewidth}
     \centering
     \includegraphics[width=1.33in, height=.8in]{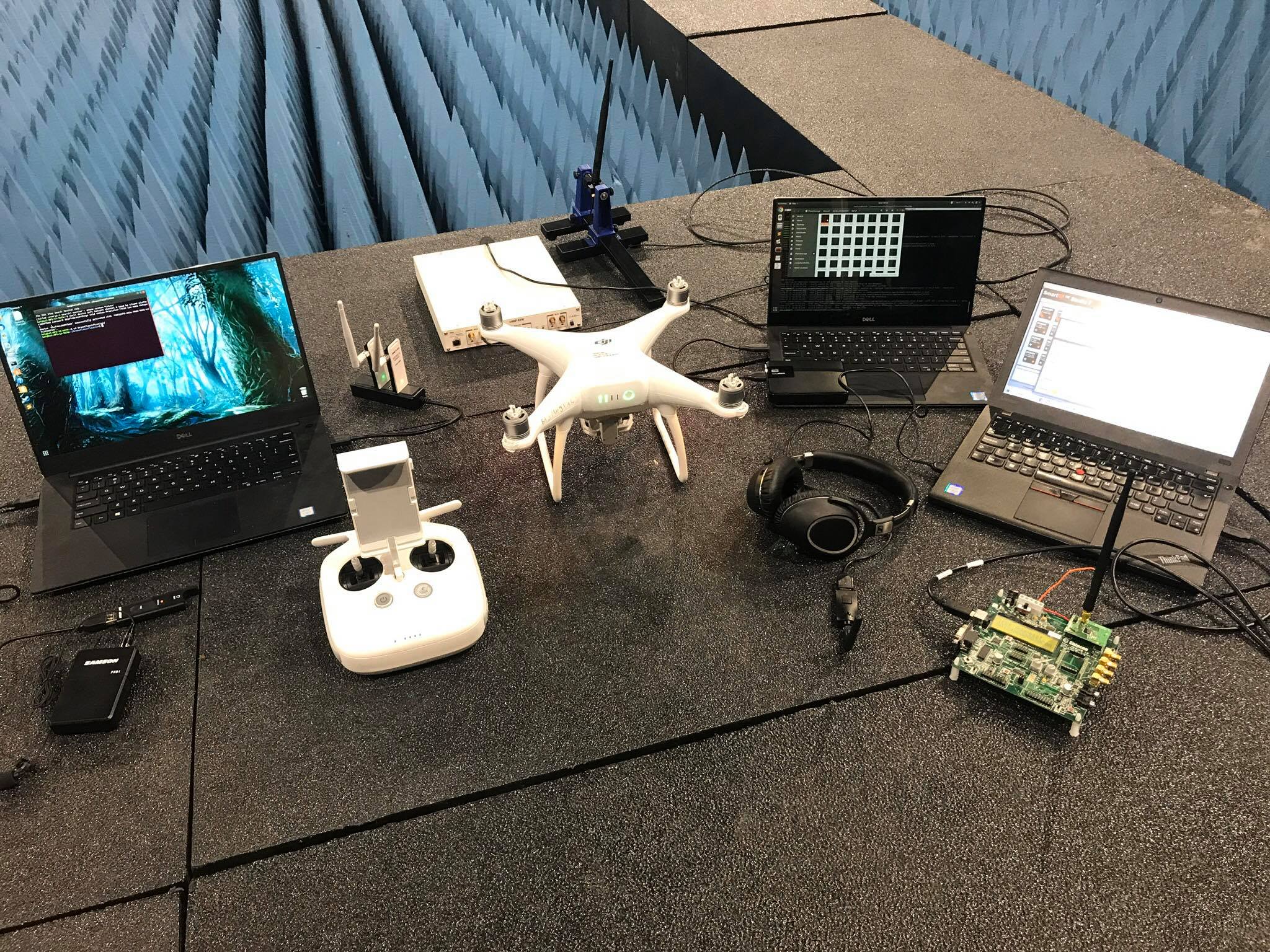}
    \caption{Devices used in \dataset{} measurements.}
    \label{fig:devices}
  \end{minipage}
\end{figure}

\subsection{\system{} System}
\bfpara{Implementation} We used an off-the-shelf Ettus USRP X310 to record the RF spectrum of 100 MHz in the 2.4GHz ISM band. The USRP is connected via 10G ethernet to a host computer equipped with a 6-core Intel Core i7-8700@3.2GHz processor, NVIDIA GeForce GTX 1080 Graphics Card, and 32 GB RAM. The integrated implementation of \system{} consists of two main parts. The first part, written in C++, is responsible for collecting RF samples from the USRP and handling the RF input compression. We implemented and optimized the RF-centric compression algorithm based on GNU Radio VOLK library. It is noted that the FFT computation is handled on the host CPU instead of the GPU. The detection module is written in Python based on the \smodelnew{} framework and utilizes the GPU \ignore{to speed up the training process}. Data communications between the RF-centric compression layer and the rest of the network are enabled by a custom message passing protocol using Google Protobuf and the ZeroMQ messaging library.

\bfpara{Real-time Microbenchmarks}
\system{} achieves real-time performance. In order to understand its limits, we benchmark each module in the pipeline. Given the monitored spectrum bandwidth of 100 MHz, a module is real-time only if its processing rate exceeds 100 Msamps/s (million samples per second). The throughput of various modules was measured on the host computer to assess the real-time capability of \system{}. It is emphasized that the RF detection module is run in parallel with the FFT and RF-centric compression modules. Based on the measurements presented in \Cref{tab:benchmark}, we observe that the bottleneck of our system is the RF detection module, yet sustaining the incoming samples rate of 100 Msamps/s. 


\bfpara{Performance in anechoic chamber} We evaluated the RF identification capabilities of \system{} with data collected in a $60 \times 60 \times 30$ ft anechoic chamber (\Cref{fig:chamber}). We created a real-life congested environment by setting up the transmissions for various RF devices (\Cref{fig:devices}) and positioning them in different locations inside the room. We recorded and evaluated on 90,426 labeled RF emissions. Due to the more complex collision patterns introduced by the congested environment that results in the increasing amount of false negatives, the detection suffers a slight decline compared to the previous test set, yet still maintain over 88 mAPs for all three IoU thresholds (96.95, 94.63 and 88.42 for $IoU_{0.25}$, $IoU_{0.5}$ and $IoU_{0.75}$ compared to 99.27, 98.70, and 92.21 in the test set in \Cref{sec:train_eval_onl}, respectively). \Cref{fig:det_perf_anec} shows examples of the RF identification, with Wi-Fi, Bluetooth, ZigBee, Lightbridge and XPD emissions identified by the green, yellow, red, blue and purple rectangular boxes, respectively. It illustrates that \system{} is able to provide accurate RF category and precise spectro-temporal position for every single RF emission under various transmission and collision patterns. 

\bfpara{Performance in the wild} We collected and labelled 1,093 emissions in an extremely congested spectrum in the wild (examples are shown in \Cref{fig:det_perf_wild}). We observed that the detection and classification difficulty is greatly increased, due to the greater volume of emissions and more complex collision patterns, which make the task challenging even for human vision. Nonetheless, our system still achieved 89.78, 85.97 and 78.89 mAP for $IoU_{0.25}$, $IoU_{0.5}$ and $IoU_{0.75}$, respectively. We reckon that the ability to detect in-the-wild emissions in real-time with just below 90 mAP for $IoU_{0.25}$ is appealing for radio systems that rely on wideband spectrum sensing to manage the communications (e.g. Wi-Fi adaptation system~\cite{wifi_adapt_mbcom13}). Besides, high-precision real-time detection (with $IoU_{0.75}$) of close to 80 mAP is promising and useful to applications that require per-emission analysis such as wireless network defense systems.

\begin{table}
\centering
\small
  \caption{Real-time microbenchmark of \system{} system.}
  \label{tab:benchmark}
  \begin{tabularx}{0.75\linewidth}{lr}
    \hline
    Module & Throughput\\
    \hline
    Detection & 130.8 Msamps/s\\
    FFT & 182.14 Msamps/s\\
    RF-centric compression & 3679.04 Msamps/s \\
  \hline
    \system{} & 130.8 Msamps/s \\
    \hline
\end{tabularx}
\end{table}

%% file: dataset.tex
\section{SPREAD\ignore{, An Open} Dataset for RFML Research}
\label{sec:dataset}
One of the goals of this work is to develop a large dataset of real-world RF communications recorded from commercial radio devices. We introduce \dataset{}\footnote{Abbreviation of \textbf{Sp}ectro-temporal \textbf{R}F \textbf{E}mission \textbf{A}nalysis \textbf{D}ataset. This dataset will be shared with the community after anonymous submission  constraint is lifted.}, an open dataset for the RFML research community. \ignore{Building \dataset{} involves the raw samples collection and preprocessing to generate 2D pictures.} In this section, we focus on describing the details of data collection, the organization and structure of \dataset{}, and various tools developed for interaction and automated tasks on the dataset.

\bfpara{Data collection}
We used the wireless devices (\Cref{tab:devices}) as RF transmitters and Software-Defined Radio running on the Ettus USRP X310 as RF recorder. We recorded emissions targeting three SNR ranges: Low (5-14 dB), Mid (15-24 dB), and High (25 dB and above). We also adjusted the transmission channels available for the Wi-Fi dongles (13 channels), the ZigBee module (16 channels) and the DJI drone (8 channels). For other devices, we created various patterns by adjusting transmission time and period. 

\ignore{When recording the wireless transmissions, we targeted three different SNR ranges: 
For Wi-Fi, we recorded randomly generated traffic over all available channels (channel 1 to 13). The Wi-Fi rate adaptation was enabled that managed the transmission rate between 30-40 Mbps. For Bluetooth, the recorded signals contained audio playback, voice transmission, file transfer and control packets, over 80 different channels. For ZigBee, we used the SmartRF Studio 7 API to control SoC CC2430EM 1.2 through the evaluation board SmartRF04EB 1.9. Random traffic of several types (data, ping, control, command) was generated on all available channels (channel 11 to 26) under continuous and periodic transmission. For Lightbridge, we transmitted real video footage from the drone DJI Phantom 4 to its controller in all available channels (channel 13 to 20), as well as the uplink Bluetooth control traffic. Lastly, we used a wireless microphone to capture over-the-air transmissions.}

\ignore{In order to minimize interference, external antennas were replaced with wired connections and an EMI protection cloth was used.} 
For the devices with external antennas such as Wi-Fi, Bluetooth, ZigBee, we recorded the RF emissions via cables connected directly to the receiver. For Lightbridge and XPD devices that lack external antennas, we used EMI-shielding fabric to eliminate unwanted external RF interference. Additionally, some of the experiments took place in an anechoic chamber for isolated and combined over-the-air transmissions under a real propagation scenario (\Cref{fig:chamber}).

\bfpara{Structure}
Besides raw data recordings, metadata and auxiliary data generated during the processing are also stored\ignore{ for reconfiguring, resuming, and speeding up the system}. \dataset{}'s top level structure is categorized into \emph{Recordings}, \emph{Pictures}, and \emph{Global configuration}. \emph{Recordings} consist of the time-domain complex samples captured into separate files. 
Every recording is described by a metadata file in JSON format containing (1) the experiment's date and duration, RF categories, channel information, center frequency, sample rate, SNR and noise power level, (2) file name and size, (3) miscellaneous dataset details such as the collecting method (recorded or from RF-based manipulation).
\emph{Pictures} consist of the original (grayscale) and compressed (RGB) RF images of the recordings with a size of $512\times{}512$. For every image there is a corresponding annotation text file, containing unordered labels for the objects found in the respective image, encoded in the format of $(class, x_c, y_c, w, h)$.
Finally, \emph{Global configuration} stores the tables of contents and global settings such as the number of FFT points, image size, compression factor. Summing up, all mentioned components construct a dataset of over 1.4 TBytes including approximately 800,000 curated, labeled RF emissions recorded using cables and EMI shielding, 90,000 emissions from anechoic chamber, and 100,000 emissions from image-based manipulations. 

\bfpara{Tools}
We provide a set of tools to manipulate the contents of the dataset. In particular, some of the available tools can be used to (1) manage the dataset (search, insert, filter, and delete elements), (2) generate synthetic RF images, (3) combine RF recordings, (4) compress pictures and annotations, and (5) automate the annotation, correction and curation processes.

        
        
        
        



        


\bfpara{Applications}
By distributing \dataset{} to the community, we share a large amount of curated and labelled RF emissions data that typically requires considerable amount of time and effort to build. We hope the dataset will spur the development of new deep learning-based RF detection and classification techniques and facilitate their training and evaluation. Several applications would significantly benefit from efficient real-time, wideband RFML such as spectrum access management (e.g., dynamic and non-cooperative spectrum sharing is preferable over crowdsourcing approach as in~\cite{crowdsensing14} in terms of spectrum coverage and performance overhead) and security (e.g., combating malicious drones). The core deep learning modules can leverage state-of-the-art computer vision approaches or fuse neural network architectures with RF techniques (processing I\&Q samples). Furthermore, sufficient RF data collection for exclusive radio technologies is made easier to achieve with the iterative learning method and supporting tools. Finally, the curated data allows to generate and insert labels for other tasks, expanding the application spectrum of the dataset.


    



\begin{figure}
    \centering
    \subcaptionbox{\label{fig:det_perf_anec}Inside anechoic chamber}{
        \includegraphics[width=.22\linewidth]{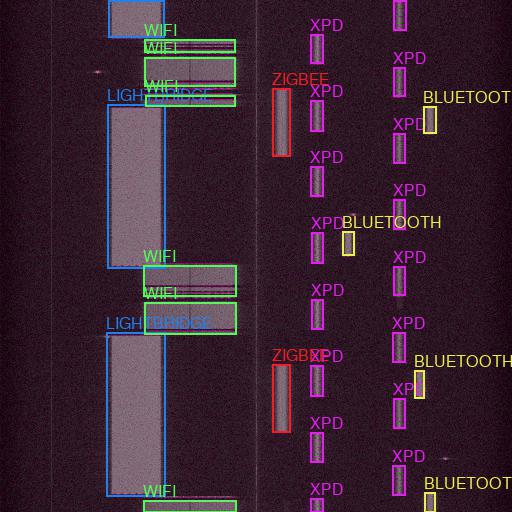}
        \includegraphics[width=.22\linewidth]{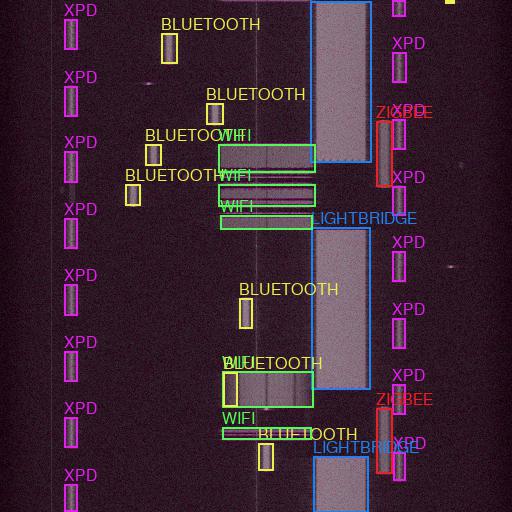}
    }
    \subcaptionbox{\label{fig:det_perf_wild}In the wild}{
        \includegraphics[width=.22\linewidth]{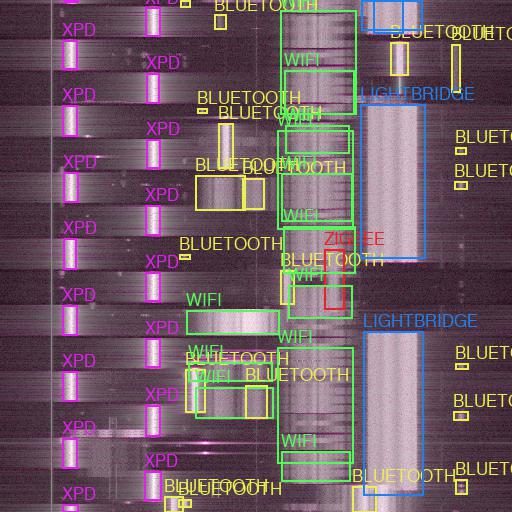}
        \includegraphics[width=.22\linewidth]{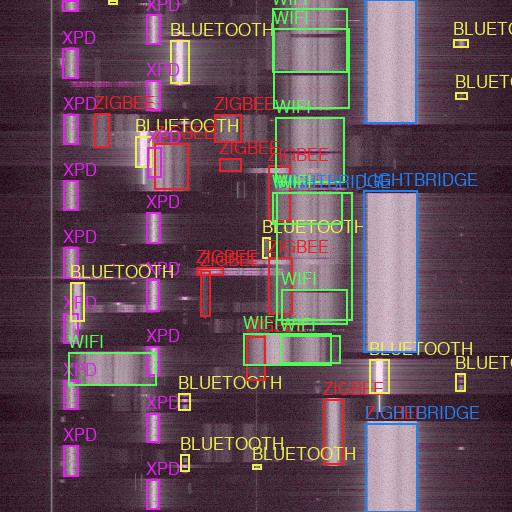}
    }   
    \caption{\system{}'s \ignore{desirable} detections \ignore{performance} in congested environments}
    \label{fig:detection_anechoic}
\end{figure}

%% file: related.tex
\section{Related work}

The problems considered in the paper project on multiple research areas. \ignore{In the following, }We discuss the related work in each area and contrast it with our approach and contributions.  

\bfpara{RF Identification} RF Identification problems (which we refer to as the sensing, detection, and classification of different RF technologies, devices, or signal properties)  attracted significant attention in the research community over the past decades. Traditional RF signal classification methods  heavily rely on extensive engineering of expert features such as higher order signal statistics (e.g., cumulants~\cite{cumulant03}, moments~\cite{moments92}), or cyclostationary signatures~\cite{cyc08}. These methods aim to identify unique observable signatures in RF transmissions. However, finding robust signatures is not easy, and requires significant effort and expert knowledge. With the ability of learning distinguished features through massive data captured, Deep Learning (DL) models have ample opportunities in such tasks, replacing conventional handcrafted features-based methods. More specifically, DL has been successfully applied to modulation classification for cognitive radios, a subclass of the identification of signal properties. \citeauthor{FGR05} proposed to use a neural network to process cyclostationary features of multiple modulated signals. Since then, several modern neural network architectures such as CNN~\cite{vtechmodrec16}, RNN~\cite{RNNmodrec} and GAN~\cite{DCGANmodrec} were used to achieve state-of-the-art performance in the recognition of different digital and analog signal modulation schemes. Despite this progress, the main drawback of previous schemes is the absence of real-time, wideband capabilities (which also results in a lack of ability to recognize simultaneous RF emissions). 

Effective wideband spectrum analysis necessitates various information about coexisting RF radio technologies instead of a single signal property as modulation type. As a matter of fact, different wireless technologies can use common modulation schemes, such as Bluetooth Low Energy (BLE) and WirelessHART which both use DSSS and FHSS. This causes the increasing interest of RF identification of different technologies in the wide, congesting, unlicensed bands. Prior work~\cite{schmidt17, bitar17} investigated RF classification for three popular 2.4 GHz technologies: IEEE 802.11 (Wi-Fi) b/g, IEEE 802.15.1 (Bluetooth), and IEEE 802.15.4 (ZigBee). They exploited existing CNN-based models, such as ResNet~\cite{resnet16} or CLDNN~\cite{cldnn2015} to classify based on either raw I\&Q samples~\cite{bitar17} or FFT-transformed samples~\cite{schmidt17}. They only processed a single signal recorded within a narrow bandwidth, and did not consider the coexistence of multiple communication technologies. Authors in~\cite{aniqua19} addressed RF classification with concurrent emissions in real-time, yet relied on dropping samples. Furthermore, supported RF categories are still limited as seen in \Cref{tab:comparison}, considering the proliferation of existing wireless and IoT technologies. In this work, we achieve significant improvements for spectro-temporal RF identification, specifically real-time, wideband processing, preserving important RF features in I\&Q data, and supporting more types of RF emissions. We believe that our iterative approach, architectures, and final models, will provide a solid basis for the critical tasks of automatic non-cooperative spectrum management mechanisms in the future, as well as other applications such as the detection of drones, and jammers.

\ignore{RF identification that targets wireless devices typically concerns about potential security and privacy issues, such as mobility and location tracking, arisen by the possibility to fingerprint devices due to unavoidable manufacturing imperfections. Expert feature-based fingerprinting methods have been investigated through various complicated tasks such as fingerprinting IEEE 802.11 transceivers \cite{VVNwifi_fp16, paradis08}, UHF RFID transponders \cite{zanetti_mobicom10}, smart phones \cite{dey2014accelprint}, or cognitive radio \cite{cog_fng17}. Besides, DL-based approaches are getting increasingly prevalent, outperforming the conventional methods, thanks to the vast volume of available Internet-of-Things (IoT) data \cite{zbee_merchant18, oracle_20}.} 


\begin{table*}
  \footnotesize
  \caption{Comparison with prior work in RF emission identification. \textmd{Our work is the first in the literature that successfully and adequately addresses the requirements for spectro-temporal RF identification, real-time and wideband processing, and sufficient real RF training data. It should be noted that we refer to the wideband processing as the ability to cover the 80 MHz ISM band. The widest bandwidth supported in the existing systems is 25 MHz \cite{aniqua19}. In this work, we are able to cover a 100 MHz bandwidth using ETTUS USRP X310.}}
  \label{tab:comparison}
  \begin{tabular}{l|c
  c
  >{\centering\arraybackslash\hspace{0pt}}m{7.5em}
  |cc|
  >{\centering\arraybackslash\hspace{0pt}}m{5em}
  c 
  >{\centering\arraybackslash\hspace{0pt}}m{5em}}
    \multirow{2}{5em}{} & \multicolumn{3}{c|}{Identification} & \multicolumn{2}{c|}{Processing} & \multicolumn{3}{c}{Training Data} \\
    & Detection & Classification & Spectro-temporal \newline Localization & Real-time & Wideband & Real RF \newline Emissions & Open Dataset & Number of \newline Technologies\\
    \hline
    This work & \boldcheckmark & \boldcheckmark & \boldcheckmark & \boldcheckmark & \boldcheckmark & \boldcheckmark & \boldcheckmark & 5\\
    Schmidt et al. \cite{schmidt17} & \boldcheckmark & \boldcheckmark & & & & & \boldcheckmark  & 3\\
    Bitar et al. \cite{bitar17} & \boldcheckmark & \boldcheckmark & & & & \boldcheckmark & & 3\\
    Baset et al. \cite{aniqua19} & \boldcheckmark & \boldcheckmark & & \boldcheckmark & & \boldcheckmark & & 3\\
    O'shea et al. \cite{oshea_yl17} & \boldcheckmark & & \boldcheckmark & & & & & N/A\\
    Lo et al. \cite{Lo_2020_WACV} & \boldcheckmark & \boldcheckmark & \boldcheckmark & & & \boldcheckmark & & 1\\
    ModRec\footnotemark \cite{vtechmodrec16, RNNmodrec, DCGANmodrec} & \boldcheckmark & \boldcheckmark & & & & \boldcheckmark & \boldcheckmark & N/A\\
    \hline
  \end{tabular}
\end{table*}

\bfpara{Deep Object Detection} In machine learning for computer vision, \citeauthor{RCNN} introduced the R-CNN object detection framework that operates in two phases: Generating \textit{proposals} of object regions (using \textit{Selective Search} algorithm) and classifying objects in those proposals (using deep CNN). This \textit{two-stages} method can generate reasonably precise detection boxes, but incurs an expensive computation cost. Following works aimed at speeding up R-CNN by improving the region proposal algorithm (the bottleneck of R-CNN) with a separately-trained network~\cite{fastRCNN} or integrating a sub-network sharing the feature maps with the detection network~\cite{fasterRCNN}. However, these techniques still lack real-time ability despite a high detection precision. When timing is the main priority, \textit{one-stage} methods such as YOLO~\cite{yolov1, yolov2, yolov3} are preferable. YOLO does not rely on excessive region proposals, but aims at predicting objects in a small set of regions obtained by gridding the image. With this approach, YOLO can effectively distinguish objects from the picture background and generate fairly accurate detections with very little computation time. Later \textit{one-stage object detection} methods such as RetinaNet~\cite{retinanet} and EfficientDet~\cite{efficientdet} tried to improve the detection accuracy with novel DL techniques including Focal Loss and Feature Pyramid Networks, at the expense of substantial loss  of prediction speed. The latest YOLOv4~\cite{yolov4} examined various novel DL techniques \ignore{preprocessing and postprocessing methods} to enhance the previous versions, and consequently achieving considerable detection improvements in the minimal trade of inference time. Based on this architecture, we designed the \smodelnew{} framework for spectro-temporal RF identification with further enhancement in prediction speed that requires milder compression, and boosting the mean Average Precision. 

\bfpara{RF Data Collection} Datasets are crucial for machine learning. \ignore{There are open datasets for modulation recognition that include both signals with \textit{simulated} channel effects and with \textit{real over-the-air}~\cite{vtechdata}} Nonetheless, large curated datasets specially for mixed RF technologies are still lacking. \citeauthor{vtechdata} dataset \cite{vtechdata} is built specially for modulation recognition task only and thus lacks of information about RF technologies. In \cite{schmidt17}, a dataset containing the recordings of Wi-Fi, Bluetooth, and Zigbee emissions was proposed and made available online. However, the recorded RF emissions were generated from a signal generator instead of commercial wireless standard devices. A recent work~\cite{parvis_magna20} discusses a larger dataset with more wireless technologies added, but it is not published or shared with the research community. More importantly, none of the mentioned datasets considers the situations of concurrent and/or colliding transmissions, or when RF emissions are observed in a wider band than their fitting bandwidth. Our dataset does not suffer from the above mentioned limitations, which hamper the advancement of DL techniques towards practical RF detection, classification, and localization. 

\footnotetext{Works for classification of signal modulation schemes.}

%% file: discussion.tex
\section{Discussion and Future Work}

\ignore{Our system has focused on spectro-temporal wideband RF identification in real-time. With the iterative learning approach, we were able to provide a large, labelled dataset for identification of wireless signals operating in the 2.4GHz ISM band.} To enhance the detection capabilities of \system{}, we plan to expand our training dataset by supporting a wider variety of RF technologies, over a wider range of frequency bands. We also plan to develop new techniques that incrementally extends the models to efficiently learn new RF emissions. Practical systems recording I\&Q samples in the wild often see unrecognizable emissions that have not been trained in advance, and misclassifying them. \ignore{Those emissions are either wrongly treated as background noise or miscategorized into different classes. To remedy this, collecting adequate data and retraining the model are required, which is time-consuming.} Because collecting adequate labelled data for retraining is time-consuming, the ability to learn and utilize unique novel RF features, is important for later development of \system{}.
\ignore{Hence, our system can be enhanced with mechanisms to separate unseen emissions from existing distinguishable ones, while efficiently learning unique novel RF features. We believe this is important for the next step of this research.}

Identifying RF emissions with very low SNRs (significantly below decodability) remains challenging because the distinguishing RF features represented in the image-based input are obscured. We intend to enhance the RF representation method by integrating unique features present in other properties of RF emissions. \ignore{For instance, divergence in phase (from I\&Q samples) may help distinguish wireless protocols.} Nonetheless, as the minimum recommended SNR for data network is around 20 dB~\cite{SNR} (which is substantially higher than \system{} threshold), very low signal strength has negligible impact on coexisting wireless communications and thus, is not an issue and not the focus of this work.
Furthermore, we wish to investigate the impact of different channel effects, such as fading, carrier frequency offset and phase offset. Finally, we intend to further push the real-time capability of the approach by constructing a fully RF-centric Deep Learning network that is less sophisticated but more effective for RF data than YOLO-based networks. \ignore{Fusion of advances in RF communications and Deep Learning will be useful for wireless identification research, as well as for other wireless security applications such as secure localization or jammer detection and countermeasures.} 

%% file: conclusion.tex
\section{Conclusion}
Understanding RF emissions in real-time is an important capability. We present \system{}, a wideband, real-time spectro-temporal RF identification system. The system provides high mean Average Precision, low latency detection, classification, and localization of RF emissions. It relies on optimized \textit{one-stage object detection} mechanisms integrated with a \textit{RF-centric compression}. Our iterative learning approach consisting of training and leveraging an \textit{offline model} and an \textit{online model} allowed us to create a curated and labeled dataset of over-the-air RF emissions. 
The deep learning models evaluation on commercial SDR peripherals proved that real-time, wideband identification is not only feasible, but also provides very high mean Average Precision. \ignore{(over 89 even for low SNR and congested spectrum in the wild)}
We also introduce \dataset{}, a large, curated, and labelled dataset that we will open to the  community for RFML research. \dataset{} spans five popular wireless technologies of samples in multiple formats amounting to 1.4 TBytes. Our iterative process developed within \system{} can be applied to new waveforms and RF emission patterns to expand the dataset. 
